\documentclass[11pt,a4paper]{article}              

\usepackage[margin=25mm]{geometry}

\usepackage{colortbl}
\usepackage{fancyhdr}
\usepackage{amssymb,amsfonts,amsmath,amsthm}
\usepackage{natbib}
\usepackage[colorlinks,citecolor=blue,urlcolor=red,bookmarks=false,hypertexnames=true]{hyperref} 
\usepackage{bstnotations}
\usepackage{graphicx}
\usepackage{subfig}
\usepackage[flushleft]{threeparttable}
\usepackage{booktabs,caption}

\usepackage{algpseudocode}
\usepackage{algorithm}
\usepackage{algorithmicx}
\newcommand*\Let[2]{\State #1 $\gets$ #2}
\algrenewcommand\algorithmicrequire{\textbf{Initialization:}}

\usepackage{authblk}
	
\usepackage{xcolor}
\definecolor{DarkGreen}{RGB}{0,100,0}   
\definecolor{DarkBlue}{RGB}{65,105,225}

\graphicspath{{./},{./Figures/}}

\usepackage{tikz}
\usetikzlibrary{shapes.geometric, arrows,positioning}

\tikzset{
	>=stealth',
	punkt/.style={
		rectangle,
		rounded corners,
		draw=black, very thick,
		text width=6.5em,
		minimum height=2em,
		text centered},
	pil/.style={
		->,
		thick,
		shorten <=2pt,
		shorten >=2pt,}
}

\tikzstyle{counter} = [ellipse, minimum width=1.5cm, minimum height=1cm, text centered, draw=black]

\tikzstyle{decision} = [diamond, minimum width=0.5cm, minimum height=0.5cm, text centered, draw=black]

\tikzstyle{Fin} = [rectangle, rounded corners, minimum width=1cm, minimum height=1cm, text centered,text width=1cm, draw=black]

\tikzstyle{line} = [draw, -latex']
\tikzstyle{arrow} = [thick,->,>=stealth]

\tikzstyle{stop} = [rectangle, rounded corners, minimum width=1cm, minimum height=1cm,text centered, draw=black]

\tikzstyle{init} = [rectangle, rounded corners, minimum width=0.5cm, minimum height=1cm, text centered,text width=8cm, draw=black]

\begin{document}
	\title{Multi-objective robust optimization using adaptive surrogate models for problems with mixed continuous-categorical parameters} 
	
	\author[1]{M. Moustapha} \author[2]{A. Galimshina} \author[2]{G. Habert}  \author[1]{B. Sudret}
	
	\affil[1]{Chair of Risk, Safety and Uncertainty Quantification,
		
		ETH Zurich, Stefano-Franscini-Platz 5, 8093 Zurich, Switzerland}
	
	\affil[1]{Chair of Sustainable Construction,
		
		ETH Zurich, Stefano-Franscini-Platz 5, 8093 Zurich, Switzerland}
	
	\date{}
	\maketitle
	
	\abstract{Explicitly accounting for uncertainties is paramount to the safety of engineering structures. Optimization which is often carried out at the early stage of the structural design offers an ideal framework for this task. When the uncertainties are mainly affecting the objective function, robust design optimization is traditionally considered. This work further assumes the existence of multiple and competing objective functions that need to be dealt with simultaneously. The optimization problem is formulated by considering quantiles of the objective functions which allows for the combination of both optimality and robustness in a single metric. By introducing the concept of common random numbers, the resulting nested optimization problem may be solved using a general-purpose solver, herein the non-dominated sorting genetic algorithm (NSGA-II). The computational cost of such an approach is however a serious hurdle to its application in real-world problems. We therefore propose a surrogate-assisted approach using Kriging as an inexpensive approximation of the associated computational model. The proposed approach consists of sequentially carrying out NSGA-II while using an adaptively built Kriging model to estimate the quantiles. Finally, the methodology is adapted to account for mixed categorical-continuous parameters as the applications involve the selection of qualitative design parameters as well. The methodology is first applied to two analytical examples showing its efficiency. The third application relates to the selection of optimal renovation scenarios of a building considering both its life cycle cost and environmental impact. It shows that when it comes to renovation, the heating system replacement should be the priority. \\[1em] 
		
		{\bf Keywords}: Robust optimization -- Multi-objective optimization -- Kriging -- NSGA-II -- Categorical variables -- Life cycle analysis
	}
	
	\maketitle
	

\section{Introduction}
Engineering systems are nowadays systematically designed for optimal efficiency in a resource-scarce environment. The optimization process is most often carried out using computational models predicting the behavior of the system for any given configuration. While these models are becoming more and more accurate thanks to algorithmic advances and the availability of low-cost computing power, the parameters required to describe the system are not always accurately known. This is due to the inherent variability of the system parameters or simply to the designer's lack of knowledge. Regardless of their nature, \emph{uncertainties} need to be accounted for in the design process as they affect the prediction made by the computational model and henceforth the design solution. Within the probabilistic framework, various techniques have been developed for optimization under uncertainties \citep{Schueller2008, Beck2015,Lelievre2016}. The optimization problem of interest in this paper comprises two components often treated separately in the literature, namely \emph{robust design optimization} and \emph{multi-objective optimization}. The first one, which directly deals with uncertainties, entails finding an optimal solution which is at the same time not sensitive to perturbations in its vicinity. This technique has been widely studied in the literature and state-of-the-art reviews can be found in \citet{Zang2005,Beyer2007}, among others. Following the pioneering sigma-to-noise ratio laid by \citet{Taguchi1989}, the most widely used metric of robustness combines the performance first- and second-order moments \citep{Doltsinis2004,Beck2015}. This naturally leads to a multi-objective optimization problem. More recently, \emph{quantiles} of the performance function have been considered \citep{Pujol2009,Razaaly2020,Sabater2021}. Owing to its simplicity and straightforward account of uncertainties through a unique scalar, this approach is considered in the present work.

The second component, \emph{i.e.}, multi-objective optimization, relates to the presence of multiple performance functions. Very often the objectives described by these functions are conflicting and a compromise needs to be found. The associated methods may be classified into \emph{a priori} and \emph{a posteriori} approaches, according to when the decision maker sets preferences among the different objectives. In the former, the designer reformulates the multi-objective problem into one or a series of single objectives by aggregation \citep{Miettinen1999,Weck2004,Ehrgott2005,Liang2017}. In contrast, in \emph{a posteriori} methods no preference is set before the optimization is carried out. Trade-off solutions are found first and only then are the decision makers given possible design solutions to choose from \citep{Marler2004}. 

Evolutionary algorithms are another class of methods that have emerged and have shown to be powerful and particularly suited for multi-objective optimization, as they possess several desirable characteristics \citep{Zitzler2004, Emmerich2018}. These are basically metaheuristics that evolve a set of candidate designs by various mechanisms so as to converge to an approximation of the so-called \emph{Pareto front}, \emph{i.e.}, a series of optimal design solutions describing a trade-off between competing objectives. Examples of such methods include the vector evaluated genetic algorithm \citep{Schaffer1985}, the multi-objective genetic algorithm \citep{Fonseca1993}, the strength Pareto evolutionary algorithm 2 (SPEA2) \citep{Zitzler1999} the Pareto archived evolution strategy \citep{Knowles2000} or the non-dominated sorting genetic algorithm II (NSGA-II) \citep{Deb2002}. The latter is considered in this work as it is efficient and easy to implement.

A commonly known drawback of evolutionary algorithms, and in particular of NSGA-II, is their high computational cost. This is even more problematic when the objective function stems from the propagation of uncertainties through a possibly expensive-to-evaluate computational model. Surrogate modelling is a proven solution to address these cost issues and has been widely used in both uncertainty quantification and design optimization fields. A surrogate model is an inexpensive approximation of a model that aims at replacing it in analyses requiring its repeated evaluation. One of the most popular methods, namely Kriging a.k.a. Gaussian process (GP) modelling \citep{Sacks1989, Santner2003, Rasmussen2006}, is considered in this work. Kriging has been extensively used for both multi-objective and robust optimization problems, as the brief literature review in Section~\ref{sec:Kriging:LitRev} shows.

Finally, the developments in this paper were motivated by an application related to the optimal selection of renovation scenarios for building renovation under uncertainties \citep{GalimshinaBE2020, GalimshinaENB2021}. The formulated optimization problem is not only multi-objective and robust but also contains a subset of parameters which are categorical. \emph{Categorical variables} are characterized by the fact that they are composed of a discrete and finite set of values which do not have any intrinsic ordering. There are also often referred to as \emph{qualitative} or \emph{nominal}. In the case of building renovation, a typical example is the heating system, which may be selected from technical solutions such as oil, gas, heat pump, etc. To address this aspect, we consider slight adaptations of both NSGA-II and Kriging which allow us to use general implementation of both methods without resorting to new developments.

Based on these premises, the goal of this paper is to develop a general purpose multi-objective algorithm for robust optimization problems using an adaptive Kriging model and capable of handling mixed categorical-continuous variables. To achieve this, we propose formulating the multi-objective and robust optimization problem using quantiles of the performance functions, which are computed by Monte Carlo simulation. The resulting optimization problem is solved using NSGA-II and the computational cost is reduced by the use of an adaptively built Gaussian process model. Focus is put on the exploration/exploitation balance of this adaptive scheme, which allows us to solve complex problems with a relatively small computational cost.

The paper is organized as follows. In Section~\ref{sec:ProbDef}, the quantile-based formulation of the optimization problem is presented considering separately the robust and multi-objective components. Section~\ref{sec:Solution} presents a nested level solution scheme which couples optimization and quantile estimation. Section~\ref{sec:Kriging} introduces Kriging to the framework following a short literature review. Section~\ref{sec:Categorical} presents the adaptations of the proposed methodology to handle mixed categorical-continuous variables. Finally, Section~\ref{sec:Applications} presents two applications: an analytical example and a real case study which deals with the optimal renovation of a building under uncertainties using life cycle assessment. 

\section{Problem formulation}\label{sec:ProbDef}
In this paper, we are interested in solving multi-objective \emph{and} robust optimization problems. Even though robust design often involves minimizing two conflicting objectives (\emph{e.g.}, a measure of performance and a measure of dispersion), these two classes of problems are most often treated separately in the literature. As briefly mentioned in the introduction, the state-of-the-art in robust optimization comprises various formulations. In this work, we are interested in minimizing conservative quantiles of the objective function given the uncertainties in the input. Mathematically, the following problem is solved:
\begin{equation}\label{eq:RobustOptim}
		\ve{d}^{\ast} = \arg \min_{\ve{d} \in \mathbb{D}}  Q_{\alpha} \prt{\mathfrak{c};\ve{X}\prt{\ve{d}}, \ve{Z}}
		\textrm{subject to:}  \quad f_j  \prt{\ve{d}}  \leq 0  \quad  \acc{j = 1, \ldots, c},
\end{equation}
where $Q_{\alpha}$, which represents the $\alpha$-quantile of the cost function $\mathfrak{c}$ and reads
\begin{equation}\label{eq:Quantile}
		Q_{\alpha} \prt{ \mathfrak{c}; \ve{X}\prt{\ve{d}},\ve{Z}} \equiv
		\inf \acc{q \in \mathbb{R} : \textrm{Prob}\bra{\mathfrak{c} \prt{\ve{X}\prt{\ve{d}},\ve{Z}} \leq q } \geq \alpha}.
\end{equation}
This quantile is minimized with respect to the design parameters $\ve{d} \in \mathbb{D} \subset{R}^{M_d}$, where $ \mathbb{D}$ denotes the design space which generally defines lower and upper bounds of the design variables. A set of $c$ constraint functions $\acc{f_j,  \, j = 1 \enum c}$ are additionally considered and are assumed to be simple and easy-to-evaluate analytical functions further restricting the design space. No other hard constraints are considered and to simplify the following developments, we will denote the subset of the design space that is feasible by $\mathbb{S} = \acc{\ve{d} \in \mathbb{D}: f_j\prt{\ve{d}} \leq 0, j = 1 \enum c}$. In contrast, the cost function is assumed to result from the evaluation of a complex, and possibly expensive-to-evaluate, computational model $\mathcal{M}$. This model takes as inputs the uncertain random parameters of the analysis, which are here split into two groups: the random variables $\ve{X} \sim f_{\ve{X} \mid \ve{d}}$ represent the variability associated to the design parameters $\ve{d}$ whereas $\ve{Z} \sim f_{\ve{Z}}$ are the so-called \emph{environmental variables} which affect the system response without necessarily being controlled by the designer. Manufacturing tolerances (design) and loading (environmental) are typical examples of the two categories of uncertain parameters in the context of structural engineering design. 

As in the applications considered in this work, engineering systems sometimes require the simultaneous optimization of multiple quantities of interest. Considering a set of $m$ cost functions denoted by $\acc{\mathcal{M}_k, \, k = 1 \enum m}$, the optimization problem of interest now becomes
\begin{align}\label{eq:RobustMultiObjOptim}
		\ve{d}^{\ast} =  \arg \min_{\ve{d} \in \mathbb{S}}  \acc{  Q_{\alpha_1} \prt{\mathcal{M}_1;\ve{X}\prt{\ve{d}}, \ve{Z}} \enum    Q_{\alpha_m} \prt{\mathcal{M}_m;\ve{X}\prt{\ve{d}}, \ve{Z}} },
\end{align}
where $\acc{\alpha_1 \enum \alpha_m} \in \bra{0,\,1}^m$ are the levels of the quantiles as introduced in Eq.~\eqref{eq:Quantile}. It is assumed here without loss of generality that the $m$ objective functions are derived from the same computational model $\mathcal{M}$.

\section{Problem solution}\label{sec:Solution}
\subsection{Nested levels solution scheme}
The direct solution of the optimization problem stated in Eq.~\eqref{eq:RobustMultiObjOptim} classically involves two nested levels. In the outer level, the design space is explored whereas in the inner one the quantiles corresponding to a given design choice are computed. Even though variance-reduction simulation techniques can be used for the computation of quantiles \citep{Glynn1996,Dong2017}, we consider here crude Monte Carlo simulation (MCS). Due to the fact that relatively large values of $\alpha$ are considered for robust design optimization, MCS may indeed provide accurate estimates of the quantiles with relatively few sample points. More specifically, we consider $\alpha_1 = \alpha_2 = ... = \alpha_m = \alpha = 0.90$ throughout this paper.

The first step in the inner level is to draw a Monte Carlo sample set for a given design choice $\ve{d}^{(i)}$:
\begin{equation}
	\mathcal{C}\prt{\ve{d}^{(i)}} = \acc{\prt{\ve{x}^{(j)},\ve{z}^{(j)}}, j = 1 \enum N},
\end{equation}
where $\ve{x}^{(j)}$ and $\ve{z}^{(j)}$ are realizations of $X \sim f_{\ve{X} \mid \ve{d}^{(i)}}$ and $Z \sim f_{\ve{Z}}$, respectively. The computational model is then evaluated on each of these points and henceforth the corresponding cost function values are obtained. After ordering the latter, the quantiles corresponding to each output are eventually empirically estimated and denoted by $q_{_k}\prt{\ve{d}^{(i)}}$.  

By their very nature, quantiles are random variables and plugging Monte Carlo estimates in Eq.~\eqref{eq:RobustOptim}~or~\eqref{eq:RobustMultiObjOptim} would lead to a stochastic optimization problem. Solving such a problem may be cumbersome, and even more so in the presence of multiple objectives. To avoid dealing with the resulting issues, the concept of \emph{common random numbers} is considered \citep{Spall2003}. This essentially translates into using the same stream of random numbers within iterations of the optimization problem. More specifically, the random variables $\acc{\ve{x}\prt{\ve{d}^{(1)}}, \ve{x}\prt{\ve{d}^{(2)}}, \ve{x}\prt{\ve{d}^{(3)}}, ...}$ generated at each iteration of the optimization procedure share the same seed. For the environmental variables, the same set of realizations of $\acc{\ve{z}^{(j)}, j= 1 \enum N}$ is used throughout the optimization. Using this trick, the mapping $\ve{d}^{(i)} \mapsto q_k \prt{\ve{d}^{(i)}}$ becomes deterministic. As such, Eq.~\eqref{eq:RobustMultiObjOptim} can be simplified into
\begin{equation}\label{eq:RMO_deter}
	\ve{d}^{\ast} =  \arg \min_{\ve{d} \in \mathbb{S}}  \acc{q_{1}\prt{\ve{d}} \enum q_{m}\prt{\ve{d}} }.
\end{equation}

This problem can then be solved using any classical multi-objective optimization solver. Such problems are however ill-defined in a sense, as there is generally no single solution that minimizes all the objective functions simultaneously. Therefore, the traditional approach is to seek for a set of solutions representing a compromise. To define such solutions, a widely-used concept is that of \emph{Pareto dominance} \citep{Miettinen1999,Ehrgott2005}.  Given two solutions $\ve{a}, \ve{b} \in \mathbb{D}$, 	$\ve{a}$ dominates $\ve{b}$ (denoted by $\ve{a} \prec \ve{b}$) if and only if the following two conditions hold:
\begin{equation*}
	\left\{ \begin{array}{ll}
		\forall k \in  \acc{1 \enum m}, \quad q_k\prt{\ve{a}} \leq q_k\prt{\ve{b}}, \\
		\exists k \in  \acc{1 \enum m}, \quad q_k\prt{\ve{a}} < q_k\prt{\ve{b}}.
	\end{array} \right.
\end{equation*}
The goal of the search is then to find solutions which are not dominated by any other feasible point. A set of such solutions is said to be \emph{Pareto optimal} and defined as follows:
\begin{equation}
	\mathcal{D}^{\ast} = \acc{\ve{d} \in \mathbb{S}:\,  \nexists\,  \ve{d}^\prime \in \mathbb{S}, \, \ve{d}^\prime \prec \ve{d}}.
\end{equation}
The \emph{Pareto front} $\mathcal{F} = \acc{\ve{q}\prt{\ve{d}} \in \mathbb{R}^m : \ve{d} \in \mathcal{D}^\ast}$ is the image of $\mathcal{D}^\ast$ in the objective space.

Typically, a discrete approximation of the Pareto front is sought. As mentioned in the introduction, we will consider in this paper an elitist evolutionary algorithm, namely the \emph{non-dominated sorting genetic algorithm II} (NSGA-II) \citep{Deb2002}. This algorithm is one of the most popular in the field of multi-objective optimization and is briefly described in Appendix~\ref{app:NSGA}.

NSGA-II is a powerful algorithm for multi-objective optimization. However, as all evolutionary algorithms, its effectiveness comes at the expense of a high computational cost,  in the order of hundreds to thousands of evaluations of the fitness function. On top of that, each evaluation of the fitness, \emph{i.e.}, the quantiles for a given design, requires thousands of evaluations of the computational model. The overall computational cost is then prohibitive, especially when the computational model is  expensive (\emph{e.g.}, results from a finite element analysis). To alleviate this burden, we consider surrogate models, more precisely Kriging, as described in the next section.

\section{Use of Kriging surrogates}\label{sec:Kriging}
\subsection{Basics of Kriging}\label{sec:Kriging:Basics}
Several types of surrogate models have been used to reduce the computational cost of multi-objective evolutionary algorithms \citep{Manriquez2016}. These surrogates are often embedded in an active learning scheme where they are iteratively updated so as to be especially accurate in areas of interest. Gaussian process modelling a.k.a. Kriging naturally lends itself to such approaches as it features a built-in error measure that can be used to sequentially improve its own accuracy. 

Kriging (a.k.a. Gaussian process modelling) is a stochastic method that considers the model to approximate as a realization of a Gaussian process indexed by $\ve{w} \in \mathbb{R}^M$ and defined by \citep{Sacks1989, Santner2003, Rasmussen2006}
\begin{equation}
	\widehat{\mathcal{M}}\prt{\ve{w}} = \ve{\beta}^T \ve{f} + Z\prt{\ve{w}},
\end{equation} 
where $\ve{\beta}^T \ve{f} = \sum_{j=1}^{p} \beta_j f_j\prt{\ve{w}}$ is a deterministic trend described here in a polynomial form (universal Kriging) and $Z\prt{\ve{w}}$ is a zero-mean stationary Gaussian process. The latter is completely defined by its auto-covariance function $\textrm{Cov}\bra{Z\prt{\ve{w}},Z\prt{\ve{w}^\prime}} = \sigma^2 R\prt{\ve{w},\ve{w}^\prime;\ve{\theta}}$, where $\sigma^2$ is the process variance, $R$ is an auto-correlation function and $\ve{\theta}$ are hyperparameters to calibrate. The auto-correlation function, which is also referred to as \emph{kernel}, encodes various assumptions about the process of interest such as its degree of regularity or smoothness. In this paper, we consider the Gaussian auto-correlation function, which is defined in its anisotropic form as
\begin{equation}
	R\prt{\ve{w},\ve{w}^\prime;\ve{\theta}} = \prod_{i=1}^{M}\exp \bra{-\frac{1}{2} \prt{\frac{w_i-w_i^\prime}{\theta_i^2}}^2}.
\end{equation}

The calibration of the Kriging model consists in training it on an experimental design (ED) \\ $\mathcal{D} = \acc{\prt{\ve{w}^{(1)}, \mathcal{M}\prt{\ve{w}^{(1)}}} \enum \prt{\ve{w}^{(n_0)}, \mathcal{M}\prt{\ve{w}^{(n_0)}}}}$ and hence finding estimates of the three sets of hyperparameters $\acc{\ve{\beta},\ve{\theta}, \sigma^2}$. This can be achieved using methods such as maximum likelihood estimation or cross-validation \citep{Santner2003,Bachoc2013b,Lataniotis2018}. Once the parameters are found, it is assumed that for any new point $\ve{w}$, $\widehat{\mathcal{M}}\prt{\ve{w}}$ follows a Gaussian distribution $\mathcal{N}\prt{\mu_{\widehat{\mathcal{M}}}, \sigma_{\widehat{\mathcal{M}}}^2 }$ defined by its mean and variance as follows:
\begin{equation}\label{eq:KRGpred}
	\begin{split}
		\mu_{\widehat{\mathcal{M}}}\prt{\ve{w}} & = \ve{f}^T\prt{\ve{w}} \widehat{\beta} + r\prt{\ve{w}} \ve{R}^{-1} \prt{\mathcal{Y} - \ve{F} \widehat{\beta}},\\
		\sigma_{\widehat{\mathcal{M}}}^2\prt{\ve{w}} & = \widehat{\sigma}^2 \prt{ 1 - \ve{r}\prt{\ve{w}}^T \ve{R}^{-1}  \ve{r}\prt{\ve{w}} + \ve{u}\prt{\ve{w}}^T \prt{\ve{F}^T \ve{R}^{-1} \ve{F}}^{-1} \ve{u}\prt{\ve{w}}},
	\end{split}
\end{equation}
where $\widehat{\ve{\beta}} = \prt{\ve{F}^T \ve{R}^{-1} \ve{F}}^{-1} \ve{F}^T \ve{R}^{-1} \mathcal{Y}$ is the generalized least-square estimate of the regression coefficients $\ve{\beta}$, $\widehat{\sigma}^2 = \frac{1}{N} \prt{\mathcal{Y} - \ve{F} \widehat{\ve{\beta}}}^T \ve{R}^{-1} \prt{\mathcal{Y} - \ve{F} \widehat{\ve{\beta}}}$ is the estimate of the process variance with $\ve{F} = \acc{f_j\prt{\ve{w}^{(i)}}, \, j = 1 \enum p, \, i = 1 \enum n_0 }$ being the Vandermonde matrix, $\ve{R}$ the correlation matrix with $R_{ij} = R\prt{\ve{w}^{(i)},\ve{w}^{(j)};\ve{\theta}}$   and $\mathcal{Y} = \acc{\mathcal{Y}^{(i)} = \mathcal{M}\prt{\ve{w}^{(i)}}, i = 1 \enum n_0}$. Furthermore, $\ve{r}\prt{\ve{w}}$ is a vector gathering the correlation between the current point $\ve{w}$ and the experimental design points and finally, $\ve{u}\prt{\ve{w}} = \ve{F}^T \ve{R}^{-1} \ve{r}\prt{\ve{w}} - \ve{f}\prt{\ve{w}}$ has been introduced for convenience.

The prediction for any point $\ve{w}$ is then given by the mean $\mu_{\widehat{\mathcal{M}}}\prt{\ve{w}}$ in Eq.~\eqref{eq:KRGpred}. On top of that, Kriging also provides a prediction variance $\sigma_{\widehat{\mathcal{M}}}^2\prt{\ve{w}}$ which is a built-in error measure that can be used to compute confidence intervals about the prediction. This variance is the main ingredient that has favored the development of adaptive methods such as Bayesian global optimization, as shown in the next section.

\subsection{Kriging for multi-objective and robust optimization}\label{sec:Kriging:LitRev}

\paragraph{Robust optimization\\}
In robust optimization, one aims at finding an optimal solution which shows little sensitivity to the various uncertainties in the system. This is generally achieved by minimizing a measure of robustness, \eg the variance and/or mean or a quantile of the cost function (See Eq.~\eqref{eq:RobustOptim}). The computation of the robustness measure coupled to the optimization incurs a prohibitive computational cost to the analysis. Hence, surrogate-assisted techniques, which are a means to reduce this cost, have received a lot of attention lately. However, due to the fact that the robustness measure that is minimized is only a by-product of the performance function which is usually approximated by a Kriging model, Bayesian optimization techniques are seldom considered. Instead, most of the literature focuses on non-adaptive schemes. In an early contribution, \citet{Chatterjee2019} performed a benchmark using several surrogate models for robust optimization. \citet{Lee2006} proposed approximating the logarithm of the variance of the performance function, which is then minimized using a simulated annealing algorithm. Some contributions however addressed the problem while considering active learning in a nested loop scheme, \emph{e.g.}, \citet{Razaaly2020} where a quantile or the mean of the response is minimized. In this work, the authors built local surrogate models for each design and used them to estimate the quantiles. Similarly, \citet{Ribaud2020} proposed to minimize the Taylor series expansions of the performance function's expectation and variance. They considered various configurations for which they used either the expected improvement (or its multi-points version) of the quantities of interest. Finally, \citet{Sabater2021} developed a method based on Bayesian quantile regression to which they added an infill sampling criterion for the minimization of the quantile.

\paragraph{Multi-objective optimization\\}
The naive and direct approach to leverage the optimization CPU cost with Kriging is to calibrate the latter using the experimental design $\mathcal{D}$ and then to use it in lieu of the original model throughout the optimization process. This approach is however expensive as it would require an accurate Kriging model in the entire design space. A more elaborate technique has been developed under the framework of \emph{efficient global optimization} (EGO, \citet{Jones1998}) where the experimental design is sequentially enriched to find the minimizer of the approximated function. This enrichment is made by maximizing the \emph{expected improvement} (EI), which is a merit function indicating how likely a point is to improve the currently observed minimum given the Kriging prediction and variance. The EGO framework has shown to be very efficient and new merit functions have been proposed to include other considerations such as a better control of the exploration/exploitation balance and the inclusion of constraints \citep{Schonlau1998,Bichon2008}. The first extension for multi-objective optimization was proposed by \citet{Knowles2006} through the ParEGO algorithm. In ParEGO, the expected improvement is applied on the scalarized objective function derived using the augmented Tchebycheff approach. By varying weights in the scalarization, a Pareto front can be found. \citet{Zhang2010} proposed a similar approach using both the Tchebycheff and weighted sum decomposition techniques with the possibility of adding multiple well-spread points simultaneously. However, such approaches inherit the pitfalls of the underlying decomposition methods. \citet{Keane2006} proposed new extensions of EI that allows one to directly find a Pareto front without transforming the multi-objective problem into a mono-objective one. Another adaptation of a mono-objective criterion has been proposed by \citet{Svenson2010} with their expected maximin improvement function. This is based on the Pareto dominance concept and involves a multi-dimensional integration problem which is solved by Monte Carlo simulation, except for the case when $m=2$ for which an analytical solution is provided. Apart from such adaptations, some researchers have proposed new improvement functions directly built for a discrete approximation of the Pareto front. This includes the contribution of \citet{Shu2020} where a new acquisition function based on a modified hypervolume improvement and modified overall spread is proposed. Moving away from the EGO paradigm, \citet{Picheny2015} proposed a stepwise uncertainty reduction (SUR) method for multi-objective optimization. Alternatively, the hypervolume measure of a set, which is the volume of the subspace dominated by the set up to a reference point, has been used to derive infill criteria. Namely, \citet{Emmerich2006} proposed a hypervolume-based improvement criterion. Originally based on Monte Carlo simulation, \citet{Emmerich2011} proposed a procedure to compute the criterion numerically. This method, however, does not scale well with the number of objective functions or samples in the approximated Pareto set. Efforts have been put in reducing this computational cost. This includes the contributions of \citet{Couckyut2014,Hupkens2015,Yang2019}. They all involve solving numerically the underlying integration problems by partitioning the space into slices using various methods. Finally, another attempt at decreasing the cost has been made by \citet{Gaudrie2020} where the authors proposed to focus the search of the Pareto front to a subspace according to some design preference. By doing so, the size of the discrete Pareto set can be reduced and the CPU cost of estimating the expected hypervolume improvement altogether.

In this work, we rather use a two-level nested approach as described in Section~\ref{sec:PropAppr}. We do not consider a direct Bayesian global optimization approach for two reasons. First, as shown above, such techniques in the context of multi-objective optimization are computationally intensive. Second, and most of all, the functions to optimize are not the direct responses of the model that would be approximated by a surrogate because of the robustness component of the problem. More specifically, in Bayesian optimization, the cost function  $\mathfrak{c}$ is approximated by the GP model and minimized at the same time. In contrast, we consider here as objective function the quantile $Q_{\alpha}\prt{\mathfrak{c}; \ve{X}\prt{\ve{d}},\ve{Z}}$. Using this quantile directly in Bayesian optimization would require approximating the mapping $\ve{d} \mapsto Q_{\alpha}\prt{\mathfrak{c}; \ve{X}\prt{\ve{d}},\ve{Z}}$. However, computing the quantile for a given value of $\ve{d}$ requires running a Monte Carlo simulation where the underlying expensive computational model $\mathcal{M}$ is repeatedly evaluated. The overall computational cost would therefore be prohibitive. For this reason, we resort to a nested two-level approach where the approximation of the computational model and the optimization problems are decoupled.

\subsection{Proposed approach}\label{sec:PropAppr}

\paragraph{Motivation\\}
In the previous section, we have very briefly reviewed the literature for multi-objective and robust optimization. The two topics were treated separately because, to the authors' best knowledge, very little to no research has been done for the combined problem. It is important to stress here that by ``multi-objective and robust optimization'' we disregard robust optimization methods that are formulated by minimizing the mean and the variance of a single performance function using a multi-objective optimization scheme. Such methods are often coined multi-objective robust optimization in the literature but are not the object of the present paper. We instead consider problems which are defined by multiple and conflicting performance functions, regardless of the robustness measure considered.

There have been some recent works combining multi-objective robust optimization as described here with Gaussian process modelling. Most notably, \citet{Zhang2019} proposed a framework for the solution of multi-objective problems under uncertainties using an adaptive Kriging model built in an augmented space. The latter is introduced as a means to combine the design and random variables space and hence to allow for the construction of a unique surrogate model that could simultaneously support optimization and uncertainty propagation \citep{Kharmanda2002,Au2005,Taflanidis2008}. Earlier works considering the augmented random space for optimization include \citet{Dubourg2011, Taflanidis2014, MoustaphaSMO2016}. 

In this work, we also consider building a surrogate model in an augmented space, however using a two-level approach instead of direct Bayesian optimization as proposed in \citet{Zhang2019}. The latter proposed to use the $\varepsilon$-constraint method to solve the multi-objective problem and devised a hybrid enrichment scheme enabled by the fact that their robustness measure is the expectation of the objective function. This is a more restrictive definition as the variance or dispersion within the performance function is not accounted for. \citet{Ribaud2020} estimated the expected improvement for the variance of the objective function by Monte Carlo simulation, which can be computationally expensive. 

In this paper, we consider a nested level approach, hence allowing us to rely on a surrogate of the objective function itself rather than that of the robustness measure, herein the quantile. The latter is then estimated using Monte Carlo simulation with the obtained surrogate and the resulting optimization problem is solved using an evolutionary algorithm, namely \emph{NSGA-II}. Coupling the optimization and the surrogate model construction in the augmented space, an enrichment scheme is devised as described in Step 6 of the algorithm described in the next section.

\paragraph{Worflow of the proposed method\\}
The workflow of the algorithm is detailed in Figure~\ref{fig:Flowchart} and summarized in the following:
\begin{figure}[!ht]
	\centering
	\includegraphics[width=0.5\textwidth]{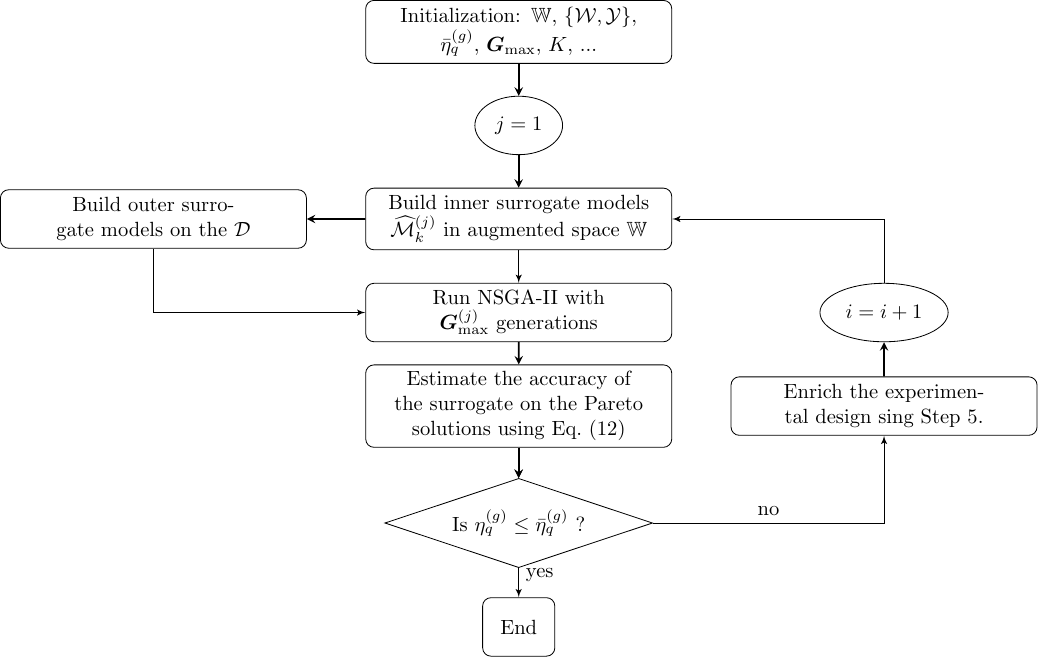}
	\caption{Flowchart of the proposed approach.}
	\label{fig:Flowchart}
\end{figure}

\begin{enumerate}
	\item  \textbf{Initialization:} The various parameters of the algorithm are initialized. This includes:
	\begin{itemize}
		\item the augmented random space, which is defined as in \citet{MoustaphaSMO2019}, \emph{i.e.}
		\begin{equation}
			\mathbb{W} = \mathbb{X} \times \mathbb{Z},
		\end{equation}
		where $\mathbb{Z}$ is the support of the random variables $\ve{Z} \sim f_{\ve{Z}}$ and $\mathbb{X}$ is the design space extended to account for the variability in the extreme values of the design variables. This \emph{confidence space}, which is defined considering the cumulative distribution of the random variables at the lower and upper design bounds, respectively denoted by $F_{X_i \mid d_i^-}$ and $F_{X_i \mid d_i^+}$, reads
		\begin{equation}
			\mathbb{X} = \prod_{i=1}^{M_d} \bra{x_i^-, \, x_i^+},
		\end{equation}
		where $x_i^{-} = F_{X_i \mid d_i^-}\prt{\alpha_{d_i^-}}$ and $x_i^{+} = F_{X_i \mid d_i^+}\prt{1 - \alpha_{d_i^+}}$ are bounds on the design random variable space with respect to confidence levels of $\alpha_{d_i^-}$ and $\alpha_{d_i^+}$. Note that if there were no uncertainties on the design parameters, we would simply use $\mathbb{X} = \mathbb{D}$;
		\item the initial experimental design $\mathcal{D} = \acc{  \prt{\mathcal{W}, \mathcal{Y}} = \prt{\ve{w}^{(i)},\ve{\mathcal{Y}}^{(i)}}}$, where the inputs $\ve{w}^{(i)} \in \mathbb{W} \subset \mathbb{R}^M$ are sampled using Latin hypercube sampling (LHS, \citet{McKay1979}) and $\ve{\mathcal{Y}}^{(i)} = \mathcal{M}\prt{\ve{w}^{(i)}} \in \mathbb{R}^m, \, i = \acc{1 \enum n_0}$, which each of the $m$ components of $\ve{\mathcal{Y}}^{(i)}$ corresponding to one of the objective functions;
		\item the NSGA-II related parameters and convergence threshold such as the maximum number of generations $G_{\textrm{max}}^{(j)}$, and
		\item the enrichment parameters, such as the number of enrichment points $K$ per iteration.
	\end{itemize}
	\item \textbf{Surrogate model construction:} $m$ Kriging models, denoted by $\widehat{\mathcal{M}}_k^{(j)}$, are built in the augmented space $\mathbb{W}$ using $\mathcal{D}$, as described in Section~\ref{sec:Kriging:Basics}. This corresponds to building a separate Kriging model for each of the $m$ performance functions.
	\item \textbf{Optimization:} The NSGA-II algorithm is then run to solve the problem in Eq.~\eqref{eq:RMO_deter} where the quantiles are computed using the surrogate model $\widehat{\mathcal{M}}^{(j)}$ in lieu of the original model. Apart from the actual convergence criteria of NSGA-II, a maximum number of generations $G_\textrm{max}^{(j)}$ is set. Its value is chosen arbitrarily low at the first iterations. The idea is that at first, emphasis is put more on the exploration of the design space. Since NSGA-II starts with a space-filling LHS over the entire design space, the first generations are still exploring. By stopping the algorithm then, we can enrich the ED by checking the accuracy of the quantiles estimated by the early sets of design samples. This allows us to direct part of the computational budget for enrichment in areas pointing towards the Pareto front without skipping some intermediate areas of interest.
	\item \textbf{Accuracy estimation:} The local accuracy of the quantiles corresponding to the points in the current Pareto front $\mathcal{F}^{(j)}$ is estimated by considering the variance of each of the $m$ Kriging models. More specifically, the relative quantile error for each objective in every point of the Pareto set is estimated as follows:
	\begin{equation}\label{eq:QRelErr}
			\eta_{q_k}^{(i)} =  \frac{q_{\alpha_k}^{+}\prt{\ve{d}^{(i)}} - q_{\alpha_k}^{-}\prt{\ve{d}^{(i)}} }{q_{\alpha_k}\prt{\ve{d}^{(i)}}}, \qquad \acc{\ve{d}^{(i)} \in \mathcal{F}^{(j)}, i = 1 \enum  \textrm{Card}\prt{\mathcal{F}^{(j)}}},
	\end{equation}
	where $q_{\alpha_k}^{\pm}$ are upper and lower bounds of the $k$-th quantile estimated using the predictors $\ve{\mu}_{\widehat{\mathcal{M}}_k^{(j)}} \pm 1.96 \, \ve{\sigma}_{\widehat{\mathcal{M}}_k^{(j)}}$. These are not actual bounds \emph{per se} but a ``kind of'' $95\%$-confidence interval indicating how locally accurate are the Kriging models. Note that when the quantiles are close to $0$, it is possible to replace the denominator in Eq.~\eqref{eq:QRelErr} by another normalizing quantity, such as the variance of the model responses at the first iteration.
	\item \textbf{Convergence check:} The convergence criterion is checked for all the points of the Pareto set with outliers filtered out to accelerate convergence. Samples with values larger than $\eta_{q_k}^{\textrm{90}} + 1.5 \, (\eta_{q_k}^{\textrm{90}} - \eta_{q_k}^{\textrm{10}})$ are considered to be outliers, where $\eta_{q_k}^{\textrm{90}}$ and $\eta_{q_k}^{\textrm{10}}$ are respectively the $90$-th and $10$-th percentile of the convergence criterion for the $k$-th objective. A usual definition of outliers is based on the interquartile range \citep{McGil1978} but we consider here a more conservative definition by rather using the interdecile range. Convergence is assumed if the relative quantile errors $\eta_{q_k}^{(i)}$ for all the points of the Pareto front (potential outliers excluded), as computed by Eq.~\eqref{eq:QRelErr}, are below a threshold $\bar{\eta}_q^{(j)}$. The latter can also be set adaptively so as to be loose in the initial exploratory cycles of the algorithm. The algorithm then goes to Step 8 if convergence is achieved or proceeds with the next step otherwise.
	\item \textbf{Enrichment:} $K$ points are added per iteration. The enrichment is carried out in two steps. The idea is to add multiple points to the ED within a single iteration by splitting them into two sets. While the first set consists of the points that directly maximize the learning function, the second set is spread out evenly among the best candidates for enrichment. In practice, they are obtained as follows:
	\begin{enumerate}
		\item First set: For each objective that does not satisfy the convergence criterion, the point in the Pareto set with the largest error is identified and denoted by $\ve{d}_k^{\textrm{max}}$. The corresponding enrichment sample is then chosen as the one which maximizes the local Kriging variance, \emph{i.e.},:
		\begin{equation}\label{eq:enr}
			\ve{w}_\textrm{next} = \arg \max_{\ve{w} \in \mathcal{C}_{q_k}} \sigma_{\widehat{\mathcal{M}}_{k}^{(j)}}\prt{\ve{w}},
		\end{equation}	
		where  $\mathcal{C}_{q_k} = \acc{\prt{\ve{x}^{(i)}\prt{\ve{d}_k^{\textrm{max}}}, \; \ve{z}^{\ve{(i)}}}, i = 1 \enum N}$ is the sample set used to compute the quantiles for the design point $\ve{d}_k^{\textrm{max}}$. The number of added points is denoted by $K_1$ and corresponds to the number of objectives for which the convergence criterion is not respected.
		\item Second set: The remaining $K_2 = K-K_1$ points are identified by first selecting all the design solutions in the Pareto front that produce errors larger than the threshold $\bar{\eta}_q^{(j)}$. This set is then reduced into $K_2$ evenly distributed points using K-means clustering. For each of these points, the corresponding enrichment sample is chosen as the one maximizing the Kriging variance among the samples used to compute the corresponding quantiles, similarly to Eq.~\eqref{eq:enr}.
	\end{enumerate} 
	\item \textbf{Experimental design update:} The original computational model is then evaluated on the $K$ enrichment points identified in the previous step, hence leading to $K$  new pairs of samples $\mathcal{D}^{(j)}_{\textrm{next}} = \acc{\prt{\ve{w}_{\textrm{next}}^{(1)}, \mathcal{M}\prt{\ve{w}_{\textrm{next}}^{(1)}}} \enum \prt{\ve{w}_{\textrm{next}}^{(K)}, \mathcal{M}\prt{\ve{w}_{\textrm{next}}^{(K)}}}}$. These samples are then added to the current experimental design, \emph{i.e.}, $\mathcal{D} \leftarrow \mathcal{D} \cup \mathcal{D}^{(j)}_{\textrm{next}}$ and the algorithm returns to Step 2.
	\item \textbf{Termination:} The outliers identified at Step 5, if any, are removed and the remaining points in the Pareto front and set are returned.
\end{enumerate}

To accelerate the whole procedure, we propose to optionally build at each iteration, and for each objective, an outer surrogate model to be used by the optimizer. The corresponding experimental design reads:
\begin{equation}
	\mathcal{D}^{(j)}_{\textrm{out}} = \acc{\prt{\ve{d}^{(i)}, \widehat{q}_{\alpha_k}\prt{\ve{d}^{(i)}}}, i = 1 \enum n_\textrm{out}^{(j)}}
\end{equation}
where $\widehat{q}_{\alpha_k}$ is the quantile estimated using the current surrogate model $\widehat{\mathcal{M}}^{(j)}_k$. Since this surrogate is built using another surrogate model, it is not necessary to use active learning at this stage. Instead, we simply draw a unique and large space-filling experimental design of size $n_\textrm{out}^{(j)}$. To increase the accuracy around the Pareto front in the outer surrogate, the experimental design inputs are updated after each cycle considering two different aspects. First, the samples in the Pareto set of the previous cycle are added to the space-filling design. Second, the accuracy of the outer surrogate w.r.t. the inner one in estimating the quantiles in the Pareto front is checked after each cycle. The related error is monitored and calculated as follows:
\begin{equation}
	\eta_{{q_k},\text{out}}^{(j)} = \max_{i \in \acc{1 \enum \text{Card}\prt{\mathcal{F}^{(j)}}}} \frac{\abs{\widehat{q}_{\alpha_k}\prt{\ve{d}^{(i)}} - \mu_{\widehat{q}_k}\prt{\ve{d}^{(i)}}} }{\widehat{q}_{\alpha_k}\prt{\ve{d}^{(i)}}},
\end{equation}
where $\mu_{\widehat{q}_k}$ is the quantile predicted by the outer surrogate model. If this error is larger than a threshold $\bar{\eta}_{q,\textrm{out}}$ the size of the ED for the outer surrogate is increased before the next cycle.

\section{The case of categorical variables}\label{sec:Categorical}
The methodology presented in the previous section assumes that all variables are continuous. However,  in some cases, and particularly in the applications considered in this paper, some of the design variables may be categorical. Categorical variables are defined in a discrete and finite space and have the particularity that they are qualitative and cannot be meaningfully ordered. As such, the Euclidean distance metric does not apply to such variables.

However, NSGA-II, Kriging and $K$-means clustering rely on such a metric since it is used in the definition of the cross-over and mutation operators, and in the evaluation of the kernel function. We will thus consider adaptations of these methods to handle the mixed categorical-continuous problems treated in this paper. The main idea of these adaptations is to remain within the general implementation of these methods, so as to use existing tools without further developments.

To highlight the nature of the variables, we introduce the notations $\ve{w} = \prt{\ve{w}_\textrm{con},\ve{w}_\textrm{cat}}$ where $\ve{w}_\textrm{con} \in \mathbb{R}^{M_\textrm{con}}$ is a vector gathering the continuous variables while $\ve{w}_\textrm{cat} \in \mathbb{R}^{M_\textrm{cat}}$ gathers the categorical parameters. Each component $w_{\textrm{cat}_j}$ can take one of $b_j$ values, called levels, and denoted by $\acc{\ell_1 \enum \ell_{b_j}}$. Hence, there is a total of $b = \prod_{j=1}^{M_\textrm{cat}} b_j$ categories. 

\paragraph{NSGA-II with mixed continuous-categorical variables\\}
Prompted by their reliance on heuristics, a large variety of both cross-over and mutation operators have been proposed in the literature throughout the years. The ones developed for NSGA-II, \emph{i.e.}, the simulated binary crossover (SBX) and polynomial mutation, are dedicated to continuous variables only (See Appendix~\ref{app:NSGA} for more details). Various adaptations have been proposed but they may only be used for discrete variables as they involve numerical rounding. For operators dedicated to categorical variables, we look into those developed for binary genetic algorithm \citep{Umbarkar2015}. Since GA allows for treating each variable separately, we first split the continuous and categorical variables into two groups. For the continuous variables, the original operators for NSGA-II are used without any modification. For the categorical variables, we consider two operators typically used in binary genetic algorithms, namely the one-point crossover operator and a simple random mutation. 

In practice, let us consider two parents whose categorical components are denoted by $\ve{w}_\textrm{cat}^{(1)}$  and $\ve{w}_\textrm{cat}^{(2)}$. One-point cross-over is achieved by randomly choosing an integer $p$ such that $1 \leq p < M_\textrm{cat} $ which will serve as a cross-over point  where the parents are split and swapped to create two offsprings, \emph{i.e.},
\begin{equation}
	\left\{
	\begin{array}{ll}
		\ve{w}_\textrm{cat}^{\textrm{offspring},(1)} = \acc{\ell_{1}^{(1)} \enum \ell_{p}^{(1)}, \ell_{p+1}^{(2)} \enum \ell_{M_\textrm{cat}}^{(2)}}, \\
		\ve{w}_\textrm{cat}^{\textrm{offspring},(2)} = \acc{\ell_{1}^{(2)} \enum \ell_{p}^{(2)}, \ell_{p+1}^{(1)} \enum \ell_{M_\textrm{cat}}^{(1)}}. \\
	\end{array}
	\right.
\end{equation}

As for the mutation, the components to be mutated are simply replaced by another level of the same variables drawn with  equi-probability. Let us assume that the component $w_{\textrm{cat},j}^{\textrm{offspring},(1)} = \ell_q$ of an offspring has been selected for mutation. It is then replaced by drawing uniformly from the set $\acc{\ell_1 \enum \ell_{M_\textrm{cat}}} \setminus \ell_q$, where each level has a probability of $1/(M_\textrm{cat}-1)$ to be selected. It should be noted at this point that mutation is only performed with a probability of $1/M$, meaning that on average only one variable (both categorical and continuous) is mutated per iteration.

\paragraph{Kriging with mixed continuous-categorical variables \\}
One of the most important features of Kriging is the auto-correlation or kernel function, which encodes assumptions about the centered Gaussian process $Z\prt{\ve{w}}$. Most Kriging applications rely on continuous variables and there is a rich list of possible kernel functions for such variables. Examples include the polynomial, Gaussian, exponential or Mat\'ern kernels \citep{Santner2003}. Such kernels are built exclusively for quantitative variables and need some tedious adaptations or pre-processing such as one-hot encoding when it comes to qualitative variables. An alternative yet cumbersome technique is to build multiple models, one associated to each category. At an intermediate level, \citet{Tran2019} proposed using a Gaussian mixture based on clusters corresponding to combination of categorical variables. A more convenient way of handling mixed categorical-continuous variable problems is through the definition of adapted kernels. This is facilitated by an important property which allows one to build valid kernels by combining other valid kernels through operators such as product, sum or ANOVA \citep{Roustant2020}. Considering the product operator and splitting the input variables into their continuous and categorical components, a kernel can be obtained as follows:
\begin{equation}
	k \prt{\ve{w},\ve{w}^\prime} = k_\textrm{con} \prt{\ve{w}_{\textrm{con}}, \ve{w}_{\textrm{con}}^\prime} \cdot k_\textrm{cat} \prt{\ve{w}_{\textrm{cat}}, \ve{w}_{\textrm{cat}}^\prime}, 
\end{equation}
where $k_\textrm{con}$ and $k_\textrm{cat}$ are kernels defined on the space of continuous and categorical variables, respectively.

Thanks to this property, it is possible to build a different kernel for each variable separately. For the continuous variables, the traditional kernels can be used without any modification. As for the categorical variables, a few approaches have been proposed in the literature. One of the earliest contributions is that of \citet{Qian2008} which however involved a tedious calibration procedure. \citet{Zhou2011} proposed an enhancement based on hypersphere decomposition with a much more simplified calibration procedure. However, they use the most generic parametrization approach which does not scale well with the overall number of categories $b$. 

More parsimonious representations have been proposed, \emph{e.g.,} the so-called \emph{compound symmetry}  which assumes the same correlation among different levels of the same categorical variable. In this work, we consider this approach combined with a simple dissimilarity measure for each categorical variable, \emph{i.e.},
	\begin{equation}
		S_{w_k w_{k}^\prime} = \left\{
		\begin{array}{ll}
			0 \quad \mbox{if} \quad w_k = w_k^\prime,\\
			1 \quad \mbox{if} \quad w_k \neq w_k^\prime; \\
		\end{array}
		\right.
	\end{equation}
	The corresponding general form of the compound symmetry kernel for one categorical variable is \citep{Pelamatti2020}:
	\begin{equation}
		k\prt{w_{\textrm{cat}_k},w_{\textrm{cat}_k}^\prime} = \left\{
		\begin{array}{ll}
			1 \quad \mbox{if} \quad w_{\textrm{cat}_k} = w_{\textrm{cat}_k}^\prime,\\
			c \quad \mbox{if} \quad w_{\textrm{cat}_k} \neq w_{\textrm{cat}_k}^\prime, \\
		\end{array}
		\right.
	\end{equation}
	where $0 < c < 1$. In this work, the same compound symmetry kernel is built but embedded within usual stationary kernels such as the Gaussian, exponential or Mat\'ern correlation functions. Considering the Gaussian kernel for instance, the corresponding uni-dimensional kernel reads:
	\begin{equation}
		k\prt{w_{\textrm{cat}_k}, w_{\textrm{cat}_k}^\prime} = \exp \left(  - \frac{1}{2}  \prt{\frac{S_{w_{\textrm{cat}_k} w_{\textrm{cat}_k}^\prime}}{\theta_k}}^2  \right) .
	\end{equation}
	Combining all dimensions, the following kernel is eventually obtained:
	\begin{equation}
			k\prt{\ve{w},\ve{w}^\prime} =  \exp  \prt{ -\frac{1}{2} \sum_{k=1}^{M_\textrm{con}} \prt{\frac{w_{\textrm{con}} - w_{\textrm{con}}^\prime}  {\theta_{\textrm{con}_k} }}^2 - \frac{1}{2} \sum_{k=1}^{M_\textrm{cat}} \prt{\frac{S_{w_{\textrm{cat}_k} w_{\textrm{cat}_k}^\prime}}{\theta_{\textrm{cat}_k}}}^2}  .	
	\end{equation}
	
	This formulation allows us to build and calibrate the Kriging model using the same tools as for the continuous case, since the hyperparameters $\theta_{\textrm{cat}_k}$ are defined similarly to $\theta_{\textrm{cont}_k}$ and both can be calibrated in the same set-up. 
	
	Finally it should be noted, as also mentioned in \citet{Roustant2020}, that despite using different constructions, this is nearly identical to the one proposed by \citet{Halstrup2016}, where the Gower distance, a distance in the mixed variable space, is embedded within the Mat\'ern auto-correlation function. The only difference is that the Euclidean distance is used for the continuous part, instead of a measure based on the range of the variables. 


This kernel is flexible enough and can be implemented easily within a generic software such as \textsc{UQLab} \citep{MarelliUQLab2014}. However, as shown in \citet{Pelamatti2020}, such kernels may be limited when there are a high number of levels or categories. In fact, a kind of stationary assumption is embedded in this construction as the kernel only depends on the difference between two levels, regardless of their actual values. More advanced techniques such as group kernels \citep{Roustant2020} or latent variables Gaussian process \citep{Zhang2020,Wang2021} have been proposed in the literature but they are not considered in this work.

Let us eventually note that the Gower distance measure is also used to compute distances in the $K$-means clustering algorithm in the enrichment scheme. In our implementation, the cluster centers are updated by calculating, in each cluster, the mean for the continuous variables and the mode for the categorical values.

\section{Applications}\label{sec:Applications}
In this section, we will consider three applications examples. The first two ones are analytical and will serve to illustrate and validate the proposed algorithm. The last one is the engineering application related to optimal building renovation strategies.

For each NSGA-II cycle, we consider a population size of $L = 100$ and a maximum number of iterations $G_{\textrm{max}} = 100$. The probability of mutation and crossover are respectively set to $0.5$ and $0.9$.

\begin{figure}[!ht]
	\centering
	\includegraphics[width=0.4\textwidth]{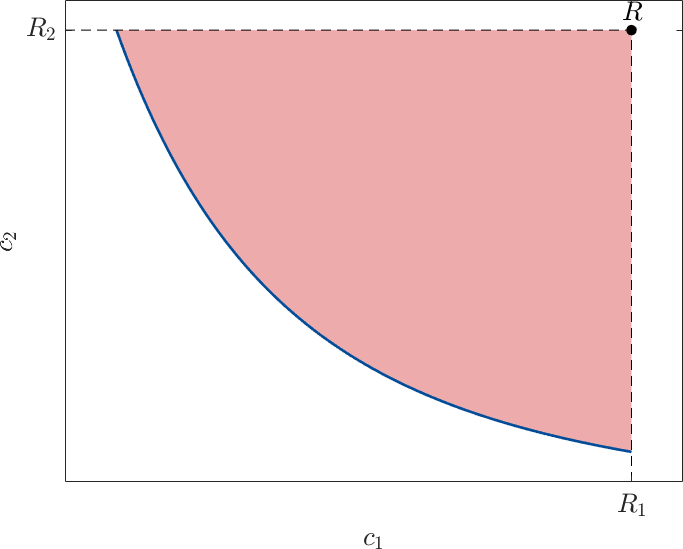}%
	\caption{Illustration of the hypervolume as the red shaded area. The reference Pareto front $ \mathcal{F}_\textrm{ref}$ is repesented by the blue line while the Nadir point $R$ is shown by the black dot.}
	\label{fig:Hypervolume}
\end{figure}

To assess the robustness of the proposed methodology, the analysis is repeated $10$ times for the first two examples. The results are summarized using boxplots, where the central mark indicates the median, the bottom and top edges of the bars indicate the 25th and 75th percentiles, respectively, and the outliers are plotted using small circles. The reference solution is obtained by solving this problem using NSGA-II and without surrogates. The accuracy is assessed by evaluating the closeness of the obtained Pareto front with the reference one. The hypervolume, which is the volume of the space dominated by the Pareto front, is chosen as basis for comparison. It is often computed up to a given reference point as illustrated in Figure~\ref{fig:Hypervolume}. In this work, the reference point is considered as the Nadir point, \emph{i.e.},
\begin{equation}
	R = \prt{R_1, R_2}= \prt{\max_{\ve{d} \in \mathcal{D}^\ast_\textrm{ref}} \mathfrak{c}_1\prt{\ve{d}}, \max_{\ve{d} \in \mathcal{D}^\ast_\textrm{ref}} \mathfrak{c}_2\prt{\ve{d}}} , 
\end{equation}
where $\mathcal{D}^\ast_\textrm{ref}$ is the Pareto set corresponding to the reference Pareto front $\mathcal{F}_\textrm{ref}$ . 

The hypervolume (here area) up to the reference point $R$ is approximated using the trapezoidal rule for integration. By denoting the one obtained from the $i$-th repetition by $A^{\prt{i}}$, the resulting relative error measure reads
\begin{equation}\label{eq:Delta_HV}
	\Delta_{\textrm{HV}}^{(i)} = \frac{\mid A^{\prt{i}}-A_\textrm{ref}\mid}{A_\textrm{ref}},
\end{equation}
where $A_\textrm{ref}$ is the hypervolume estimated using the reference solution. 

To estimate the part of the error due to the outer surrogate model, we compute the same error again, but this time using the original model and the Pareto set. Denoting the corresponding hypervolume by $A^\prime$, the following relative error is estimated: 
\begin{equation}\label{eq:Delta_HVprime}
	\Delta_{\textrm{HV}}^{\prime (i)} = \frac{\mid A^{\prime \prt{i}}-A_\textrm{ref}\mid}{A_\textrm{ref}}.
\end{equation}

\subsection{Example 1: $7$-dimensional analytical problem}
\begin{figure}[!ht]
	\centering
	\subfloat[Relative error w.r.t. surrogate model]{\label{fig:Example1:Delta:a}\includegraphics[width=0.45\textwidth]{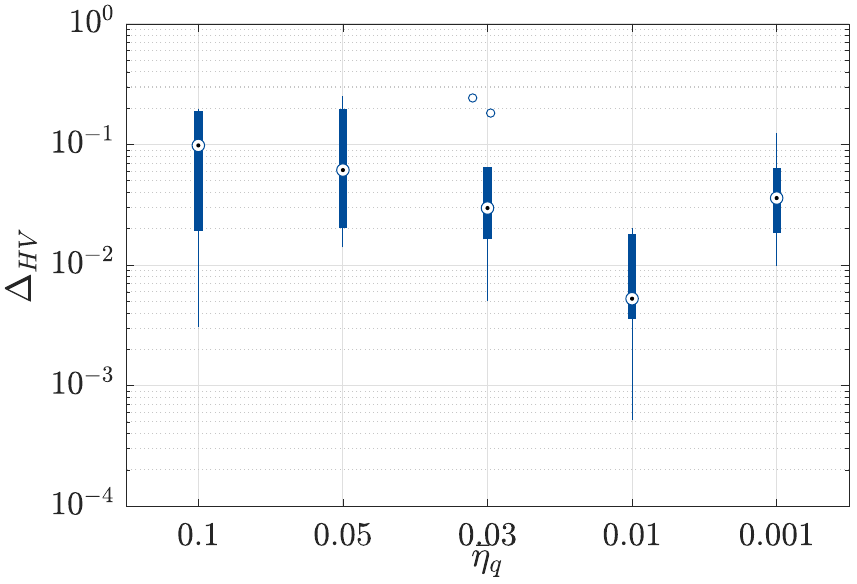}}%
	\subfloat[Relative error w.r.t. original
	model]{\label{fig:Example1:Delta:b}\includegraphics[width=0.45\textwidth]{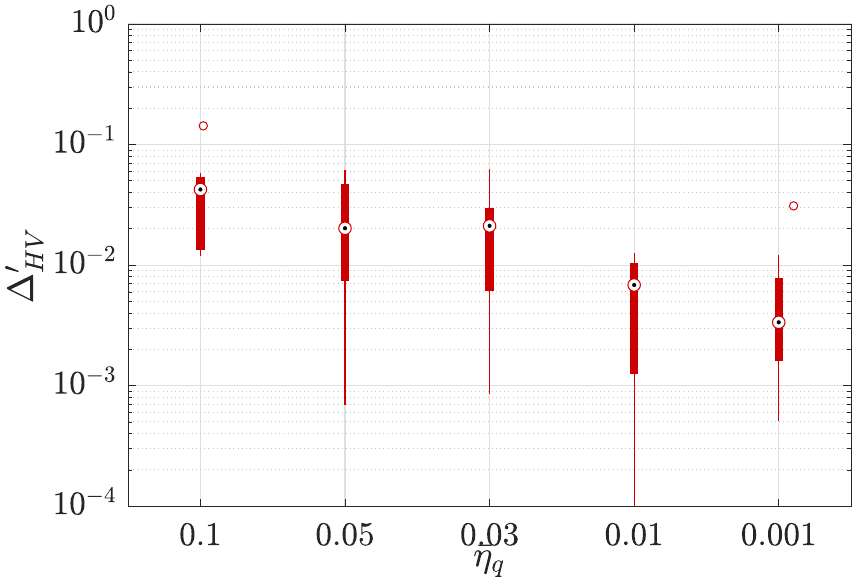}}%
	\caption{Example 1: Relative errors w.r.t. the reference hypervolume for various thresholds of the stopping criterion.}
	\label{fig:Example1:Delta}
\end{figure}

\begin{figure}[!ht]
	\centering
	\includegraphics[width=0.45\textwidth]{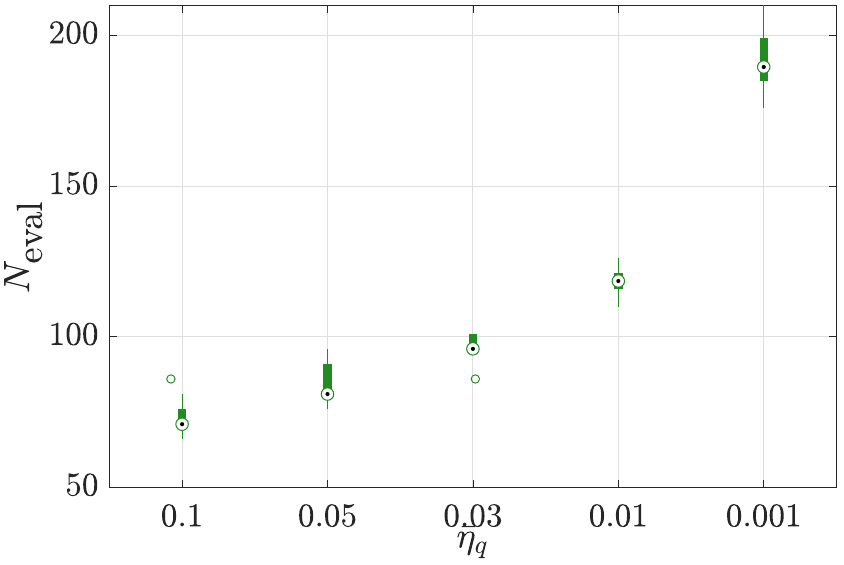}%
	\caption{Example 1: Number of model evaluations for various thresholds of the stopping criterion.}
	\label{fig:Example1:Neval}
\end{figure}

The first example considered here is an analytical problem built for illustration and validation purposes. The original problem, referred to as BNH, is popular in benchmarks for multi-objective optimization and was introduced in \citet{Binh1997}. It is a two-dimensional problem which is however deterministic and only contains continuous variables. We therefore add two categorical variables and three random variables so as to build a multi-objective and robust optimization problem. The original problem, which reads 
\begin{equation}
	\begin{split}
		\ve{d}^\ast = & \arg \min_{\ve{d} \in \bra{0, \, 5} \times \bra{0, \, 3}} \acc{\tilde{\mathfrak{c}}_1 = 4 \prt{d_{1}^{2} + d_{2}^{2}} \, ; \, \tilde{\mathfrak{c}}_2 = {\prt{d_1 - 5}}^2 + \prt{d_2 - 5}^2 } \\
		\text{subjet to:} & \acc{ \prt{d_1-5}^2 + d_{1}^{2} - 25 \leq 0 \, ; \, - \prt{d_1-8}^2 - \prt{d_2+3}^2 + 7.7 \leq 0},
	\end{split}
\end{equation} 
is modified using the following two steps:
\begin{itemize}
	\item First, the two objective functions are modified to include two categorical variables  $d_3$ and $d_4$, which can take each three possible levels:
	\begin{equation}
		\begin{split}
			\left\{ \begin{array}{lll}
				\bar{\mathfrak{c}}_1 = \tilde{\mathfrak{c}}_1 + 5  & \bar{\mathfrak{c}}_2 = \tilde{\mathfrak{c}}_2 + 5 & \textrm{if} \quad d_3 = 1,\\
				\bar{\mathfrak{c}}_1 = \tilde{\mathfrak{c}}_1 - 2 & \bar{\mathfrak{c}}_2 = \tilde{\mathfrak{c}}_2 - 2&  \textrm{if} \quad  d_3 = 2, \\
				\bar{\mathfrak{c}}_1 = \tilde{\mathfrak{c}}_1  & \bar{\mathfrak{c}}_2 = \tilde{\mathfrak{c}}_2 & \textrm{if} \quad  d_3 = 3,
			\end{array} \right.
		\end{split}
	\end{equation}
	\begin{equation}
		\begin{split}
			\left\{ \begin{array}{lll}
				\hat{\mathfrak{c}}_1 = 2 \, \bar{\mathfrak{c}}_1  & \hat{\mathfrak{c}}_2 = 2 \, \bar{\mathfrak{c}}_2 & \textrm{if} \quad d_4 = 1,\\
				\hat{\mathfrak{c}}_1 = 0.8 \, \bar{\mathfrak{c}}_1  & \hat{\mathfrak{c}}_2 = 0.95 \, \bar{\mathfrak{c}}_2 & \textrm{if} \quad d_4 = 2,\\
				\hat{\mathfrak{c}}_1 = 0.95 \, \bar{\mathfrak{c}}_1  & \hat{\mathfrak{c}}_2 = 0.8 \, \bar{\mathfrak{c}}_2 & \textrm{if} \quad d_4 = 3,\\
			\end{array} \right.
		\end{split}
	\end{equation}
	\item Then the random variables are added as follows: 
	\begin{equation}
		\begin{split}
			\left\{ \begin{array}{ll}
				\mathfrak{c}_1 = (\hat{\mathfrak{c}}_1 + z_5^2) \, z_7,\\
				\mathfrak{c}_2 = (\hat{\mathfrak{c}}_2 + z_6^2) \, z_7,
			\end{array} \right.
		\end{split}
	\end{equation}
	where $Z_5 \sim \textrm{Lognormal}(5, \, 0.5^2)$, $Z_6 \sim \textrm{Lognormal}(4, \, 0.4^2)$ and $Z_7 \sim \textrm{Gumbel}(1, \, 0.2^2)$. In this notation, the two parameters are the mean and variance of $Z_5$, $Z_6$ and $Z_7$, respectively.
\end{itemize}
The final optimization problem therefore reads:
\begin{equation}\label{eq:Example1}
	\begin{split}
		\ve{d}^{\ast} = & \arg \min_{\ve{d} \in \bra{0, \, 5} \times \bra{0, \, 3}}  \acc{Q_{\alpha} \prt{\mathfrak{c}_1;\ve{X}\prt{\ve{d}}, \ve{Z}}, \, Q_{\alpha} \prt{\mathfrak{c}_2; \ve{X}\prt{\ve{d}}, \ve{Z}} } \\
		\text{subject to:} & \acc{ \prt{d_1-5}^2 + d_{1}^{2} - 25 \leq 0 \, ; \, - \prt{d_1-8}^2 - \prt{d_2+3}^2 + 7.7 \leq 0}
	\end{split}
\end{equation}
For each of the $10$ repetitions, the analysis is started with an initial experimental design of size $n_0 = 3 M = 21$. We consider five values of the stopping criterion, namely $\bar{\eta}_q = \acc{0.1, \, 0.05, \, 0.03, \, 0.01, \, 0.001}$ (See Eq.~\eqref{eq:QRelErr}). Figure~\ref{fig:Example1:Delta} shows the relative error on the hypervolumes (as in Eq.~\eqref{eq:Delta_HV}~and~\eqref{eq:Delta_HVprime}) for each of these cases. As expected, the accuracy of the results increases as the convergence criterion is made tighter. This is however true only up to a given point. With $\bar{\eta}_q = 10^{-3}$, the criterion $\Delta_{\textrm{HV}}^{\prime}$, which is based on the Pareto set keeps decreasing while it increases noticeably when considering $\Delta_{\textrm{HV}}$ which is directly computed using the estimated Pareto front. This discrepancy can be attributed to the outer surogate model that becomes less accurate, probably due to overfitting, as the number of samples sizes is increased. It should be reminded here that the outer surrogate is built using a unique experimental design whose size may increase together with the iterations. Finally, we note that the number of evaluations of the original model increases rapidly together with the threshold for convergence, as shown in Figure~\ref{fig:Example1:Neval}.

\begin{figure}[!ht]
	\centering
	\subfloat[Convergence for $\mathfrak{c}_1$]{\label{fig:Example1:Conv:a}\includegraphics[width=0.45\textwidth]{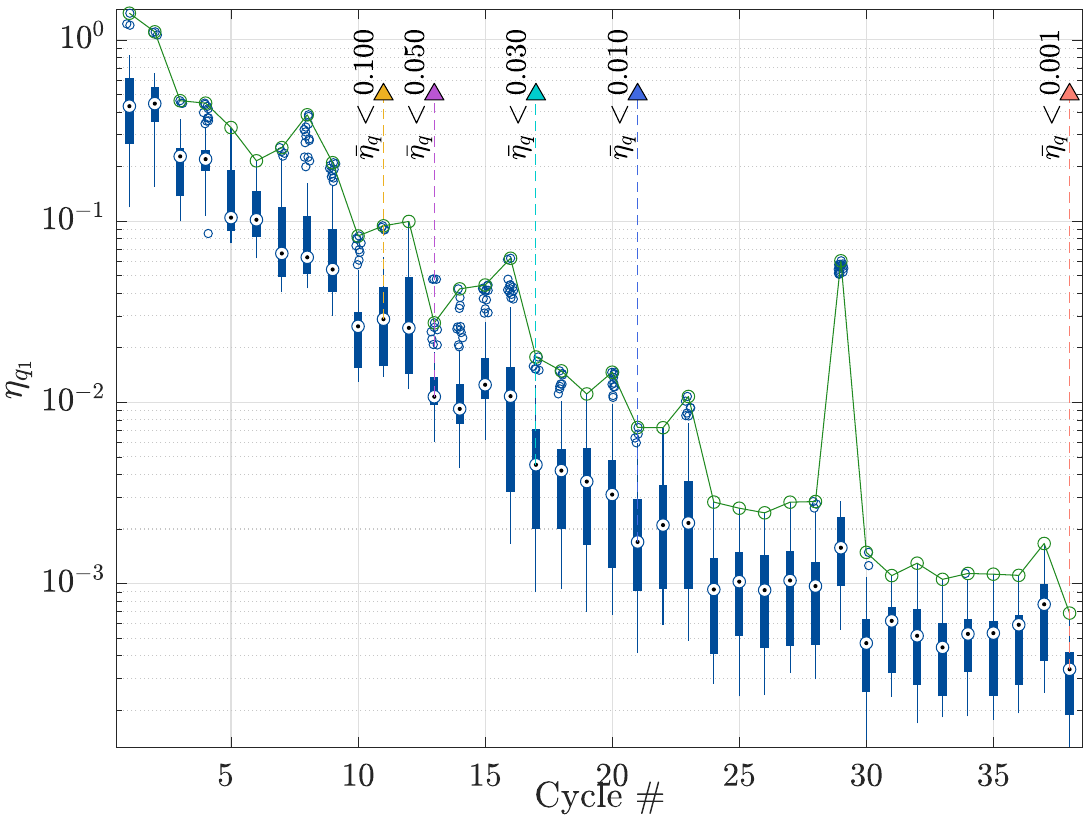}}%
	\subfloat[Convergence for $\mathfrak{c}_2$]{\label{fig:Example1:Conv:b}\includegraphics[width=0.45\textwidth]{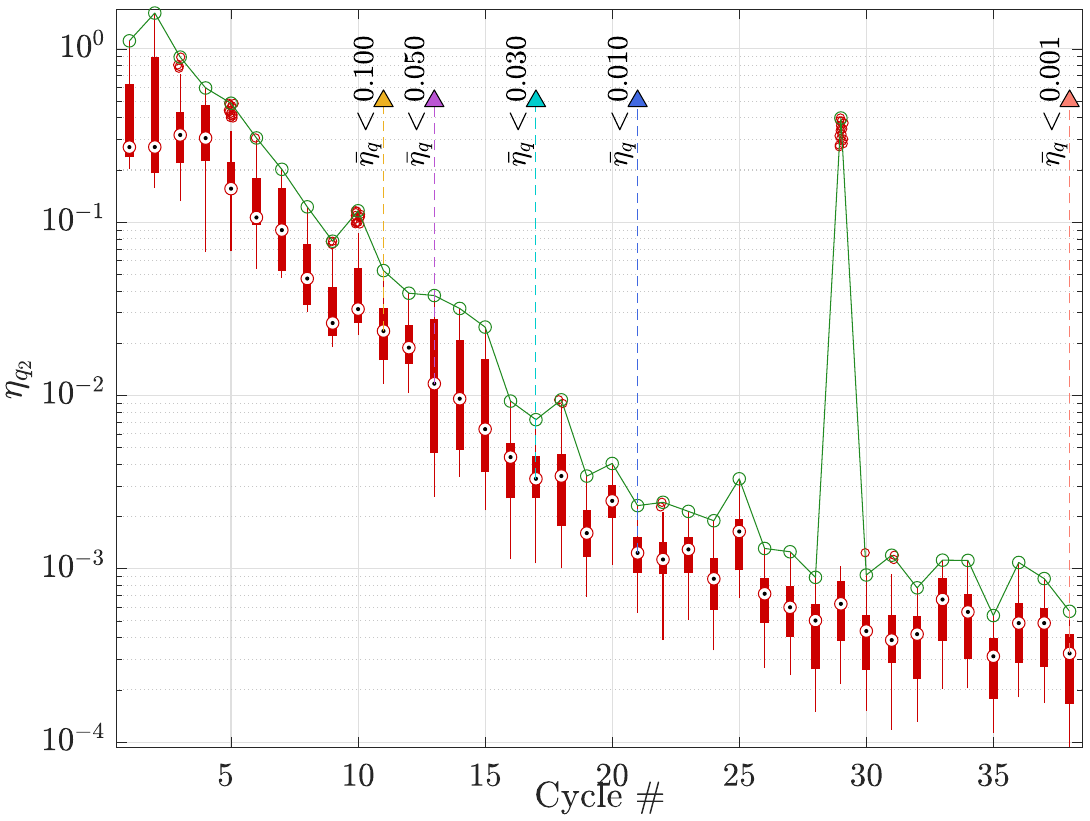}}%
	\caption{Example 1: Relative error of the $90\%$ quantiles of the costs $\mathfrak{c}_1$ and $\mathfrak{c}_2$ for the entire Pareto front at the end of each cycle. The upper convergence limit is shown by the continuous line.}
	\label{fig:Example1:Conv}
\end{figure}

To further explore these results, we consider the run with the median accuracy at the threshold of $\bar{\eta}_q = 0.03$. Figure~\ref{fig:Example1:Conv} shows the convergence of the selected analysis with the boxplots representing the relative error on the quantile for each point of the Pareto fronts. The green lines correspond to the worst quantile relative error after excluding outliers. The vertical dashed lines together with the triangular markers show where the algorithm would stop for various thresholds of the convergence criterion. After $20$ iterations, the rate of convergence decreases considerably, which explains the large added number of model evaluations required to reach the target of $\bar{\eta}_q = 10^{-3}$.

\begin{figure}[!ht]
	\centering
	\subfloat[Pareto front ]{\label{fig:Example1:Pareto:a}\includegraphics[width=0.45\textwidth]{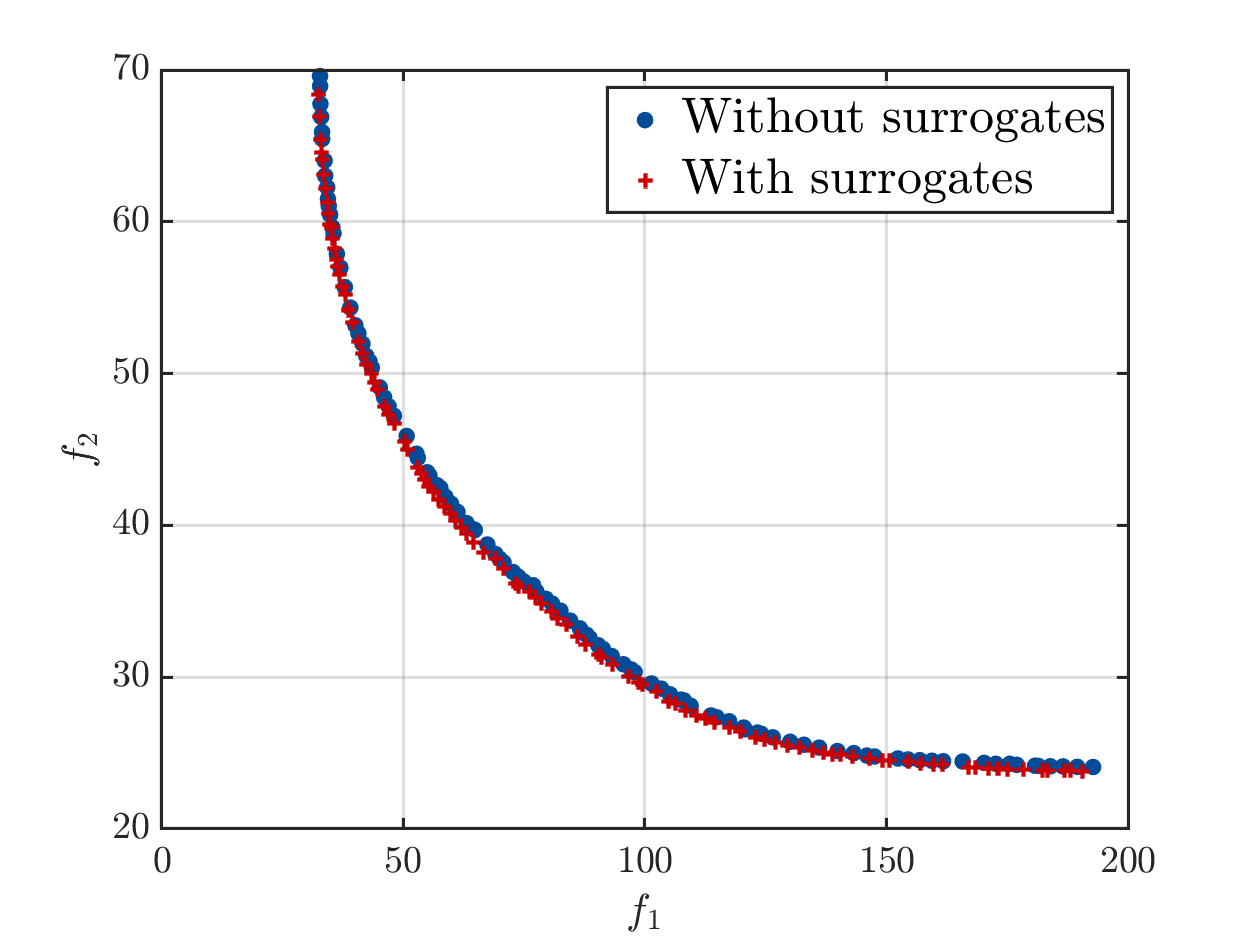}}%
	\subfloat[Pareto set]{\label{fig:Example1:Pareto:b}\includegraphics[width=0.45\textwidth]{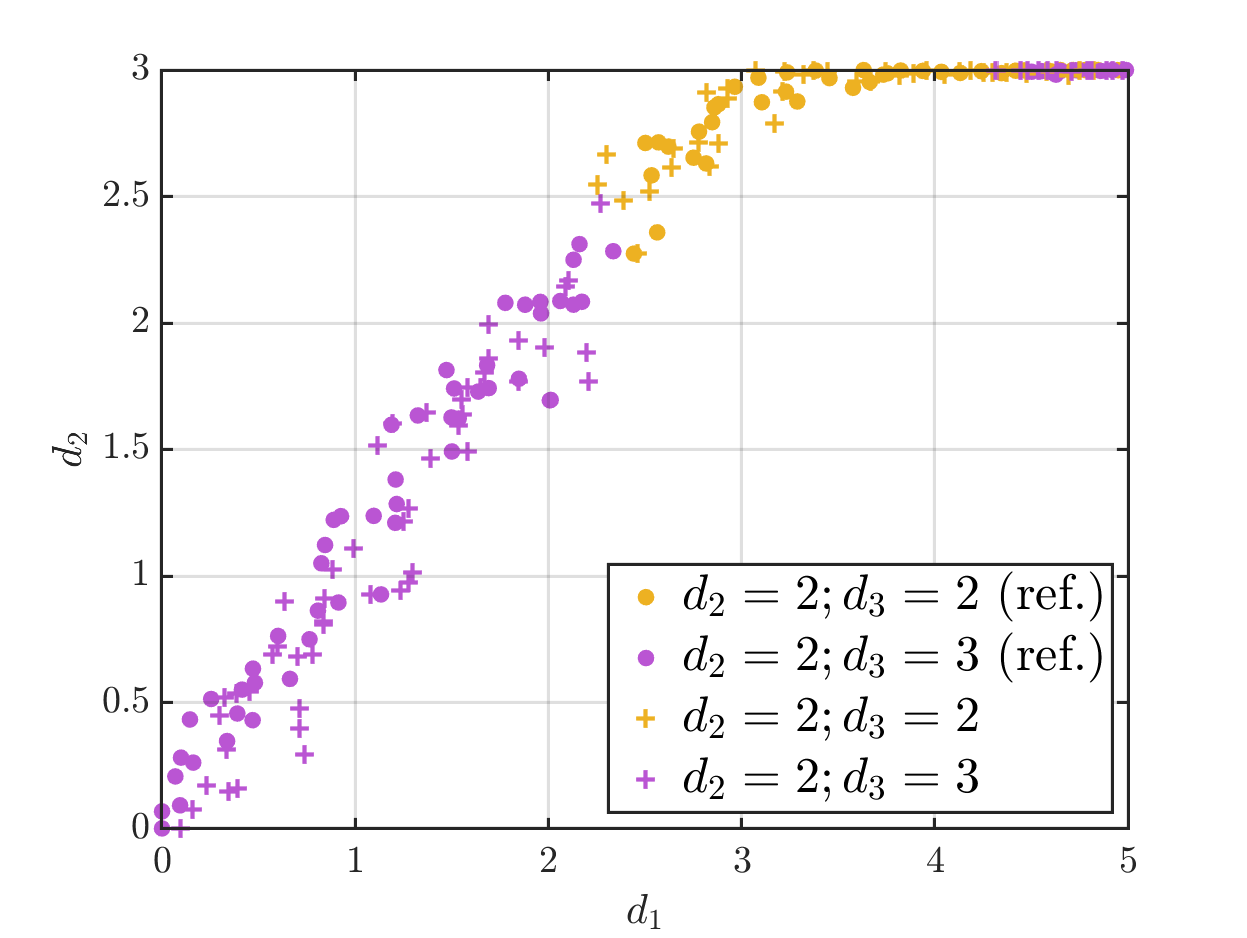}}%
	\caption{Example 1: Comparison of the Pareto fronts and sets for the results with median relative hypervolume error and the reference solution.}
	\label{fig:Example1:Pareto}
\end{figure}

Figure~\ref{fig:Example1:Pareto} shows the corresponding Pareto front (Figure~\ref{fig:Example1:Pareto:a}) and set (Figure~\ref{fig:Example1:Pareto:b}) together with the reference ones. As expected, given the way the categorical variables were introduced, there are two combinations of the categorical variables in the Pareto set: $\prt{d_3, d_4} = \prt{2, \,2}$ or $\prt{2, \,3}$. The two fronts cover the same volume and are spread in a similar way. In the input space, we can also see that the solutions cover roughly the same area. For this highlighted example, convergence is achieved in $16$ cycles with a total of $101$ evaluations of the original model. This contrasts with the total of $5 \cdot 10^7$ model evaluations ($100$ iterations of NSGA-II with $100$ samples per generation and $5,000$ samples for the estimation of the quantile for each design) required for the reference solution using a brute force approach.

\subsection{Example 2: Analytical problem with discontinuous Pareto front}
This second analytical example is adapted from \citet{Manson2021}. The Pareto front for this example is concave and presents two discontinuities. Further it only features design variables (\emph{i.e.}, there is no environmental variables), some of which are random. This allows us to showcase the versatility of the proposed algorithm. The two deterministic cost functions read
\begin{equation}
	\begin{split}
		\mathfrak{c}_1 = &
		\left\{ \begin{array}{ll}
			1 - \exp\prt{- \sum_{i=1}^{2} \prt{d_i - 1/\sqrt{2}}^2} & \textrm{if} \quad d_3 = 1,\\
			1.25 - \exp\prt{- \sum_{i=1}^{2} \prt{d_i - 1/\sqrt{2}}^2} & \textrm{if} \quad d_3 = 2,\\
		\end{array} \right. \\
		\mathfrak{c}_2 = &
		\left\{ \begin{array}{ll}
			1 - \exp\prt{- \sum_{i=1}^{2} \prt{d_i + 1/\sqrt{2}}^2} & \textrm{if} \quad d_3 = 1,\\
			0.75 - \exp\prt{- \sum_{i=1}^{2} \prt{d_i + 1/\sqrt{2}}^2} & \textrm{if} \quad d_3 = 2.\\
		\end{array} \right.
	\end{split}
\end{equation}
We then add some random variables which we associate to the design variables $d_1$ and $d_2$. Both are assumed normal, \ie{} $X_i \sim \mathcal{N}\prt{d_i,0.1^2}, \, i = \acc{1,2}$.

\begin{figure}[!ht]
	\centering
	\subfloat[Relative error w.r.t. surrogate model]{\label{fig:Example2:Delta:a}\includegraphics[width=0.45\textwidth]{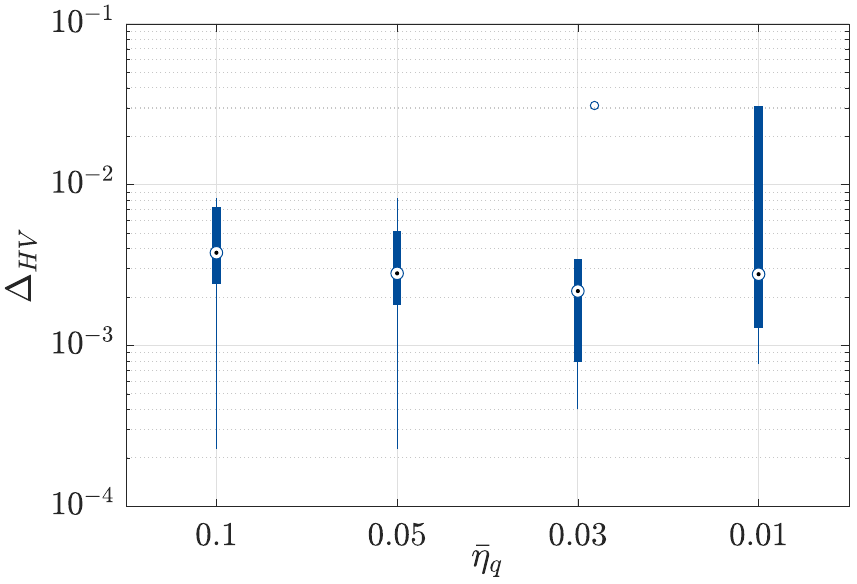}}%
	\subfloat[Relative error w.r.t. original model]{\label{fig:Example2:Delta:b}\includegraphics[width=0.45\textwidth]{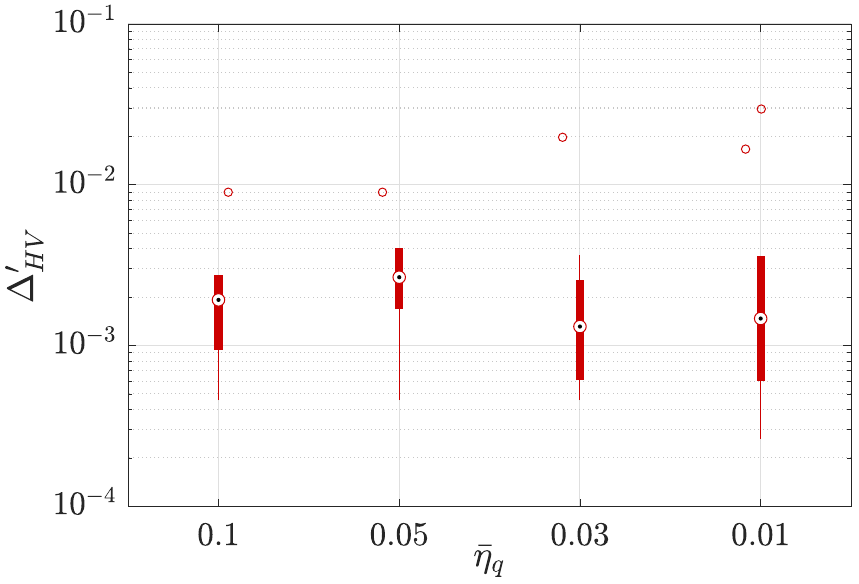}}%
	\caption{Example 2: Relative errors w.r.t. the reference hypervolume for various thresholds of the stopping criterion.}
	\label{fig:Example2:Delta}
\end{figure}

\begin{figure}[!ht]
	\centering
	\subfloat[Convergence for $\mathfrak{c}_1$]{\label{fig:Example2:Conv:a}\includegraphics[width=0.45\textwidth]{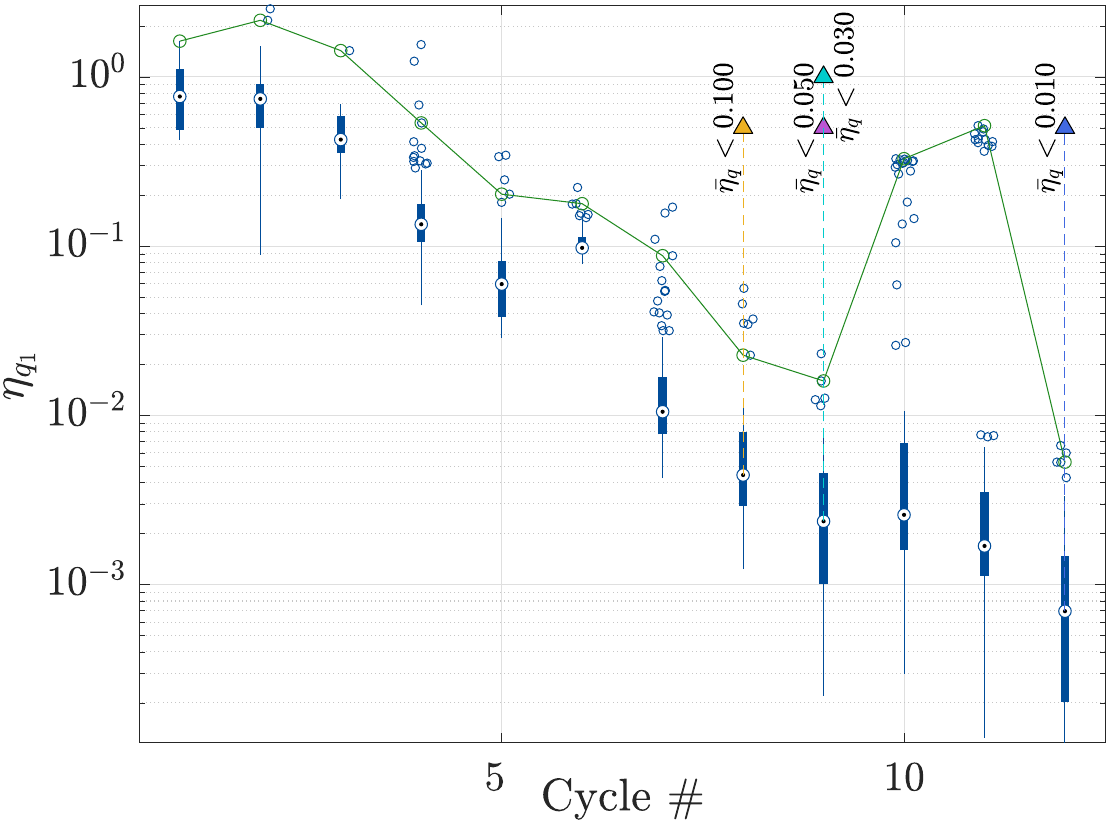}}%
	\subfloat[Convergence for $\mathfrak{c}_2$]{\label{fig:Example2:Conv:b}\includegraphics[width=0.45\textwidth]{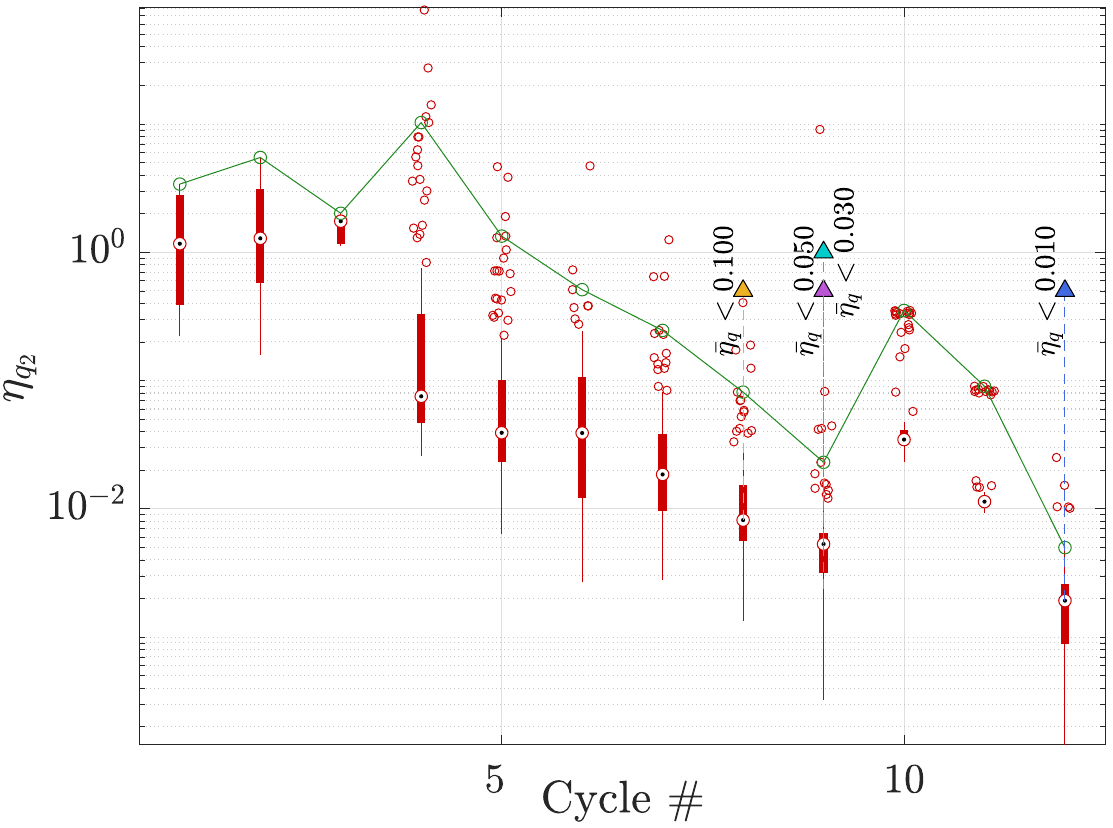}}%
	\caption{Example 2: Relative error of the quantiles of the costs $\mathfrak{c}_1$ and $\mathfrak{c}_2$ for the entire Pareto front at the end of each cycle. The upper convergence limit is shown by the continuous line.}
	\label{fig:Example2:Conv}
\end{figure}

The initial experimental design is set to $n_0 = 3\,M = 9$. For this example, the variations in accuracy due to tighter convergence criteria are not so noticeable, as can be seen in Figure~\ref{fig:Example2:Delta}. In fact, the resulting Pareto front is already accurate enough with $\bar{\eta}_q = 0.1$. It should be noted that some of the variability in estimating the relative error is due to the approximations inherent to computing the hypervolume using the trapezoidal rule for integration and the limited set of points it relies upon.
\begin{figure}[!ht]
	\centering
	\includegraphics[width=0.5\textwidth]{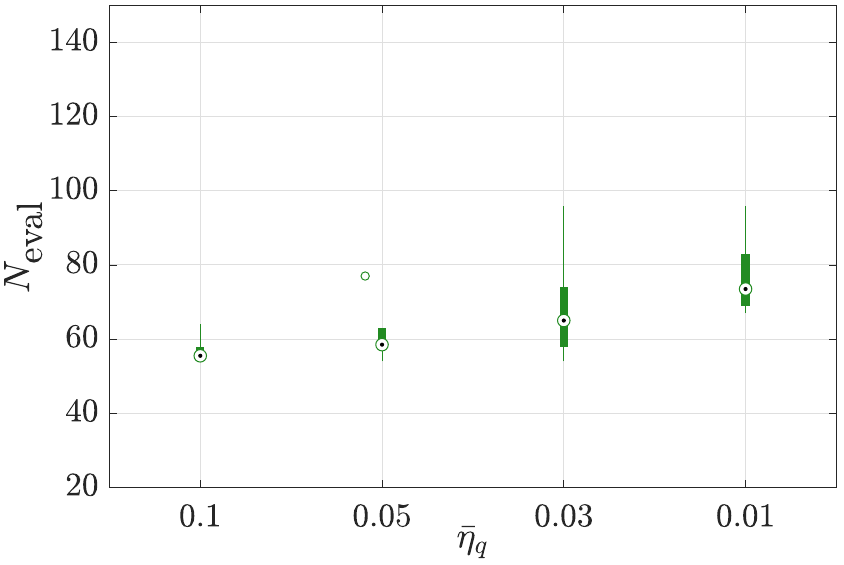}%
	\caption{Example 2: Number of model evaluations for various thresholds of the stopping criterion.}
	\label{fig:Example2:Neval}
\end{figure}
This rapid convergence may also be seen on the small increase in number of model evaluations and cycles to convergence as shown in Figures~\ref{fig:Example2:Neval}~and~\ref{fig:Example2:Conv}.

\begin{figure}[!ht]
	\centering
	\subfloat[Pareto front ]{\label{fig:Example2:Pareto:a}\includegraphics[width=0.45\textwidth]{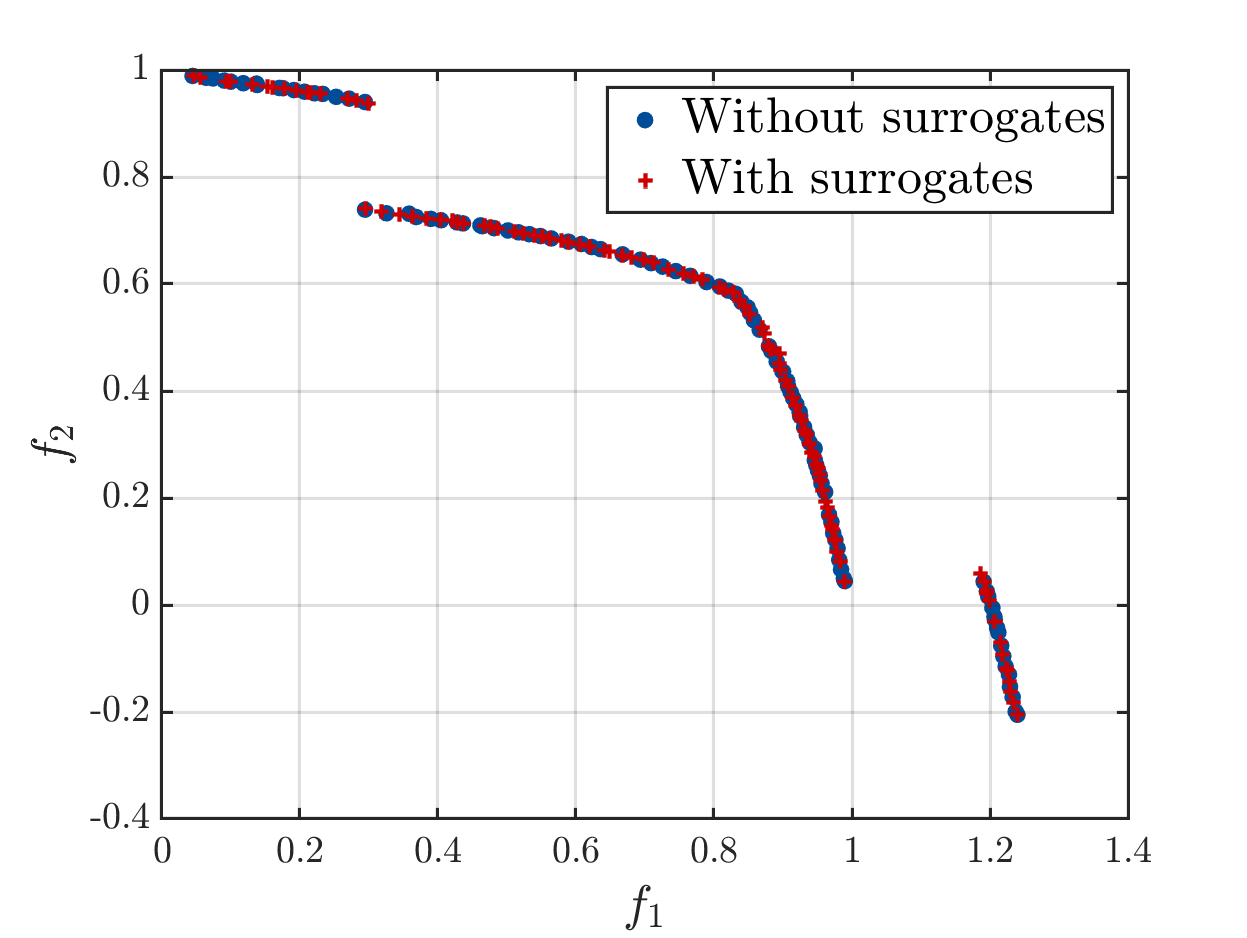}}%
	\subfloat[Pareto set]{\label{fig:Example2:Pareto:b}\includegraphics[width=0.45\textwidth]{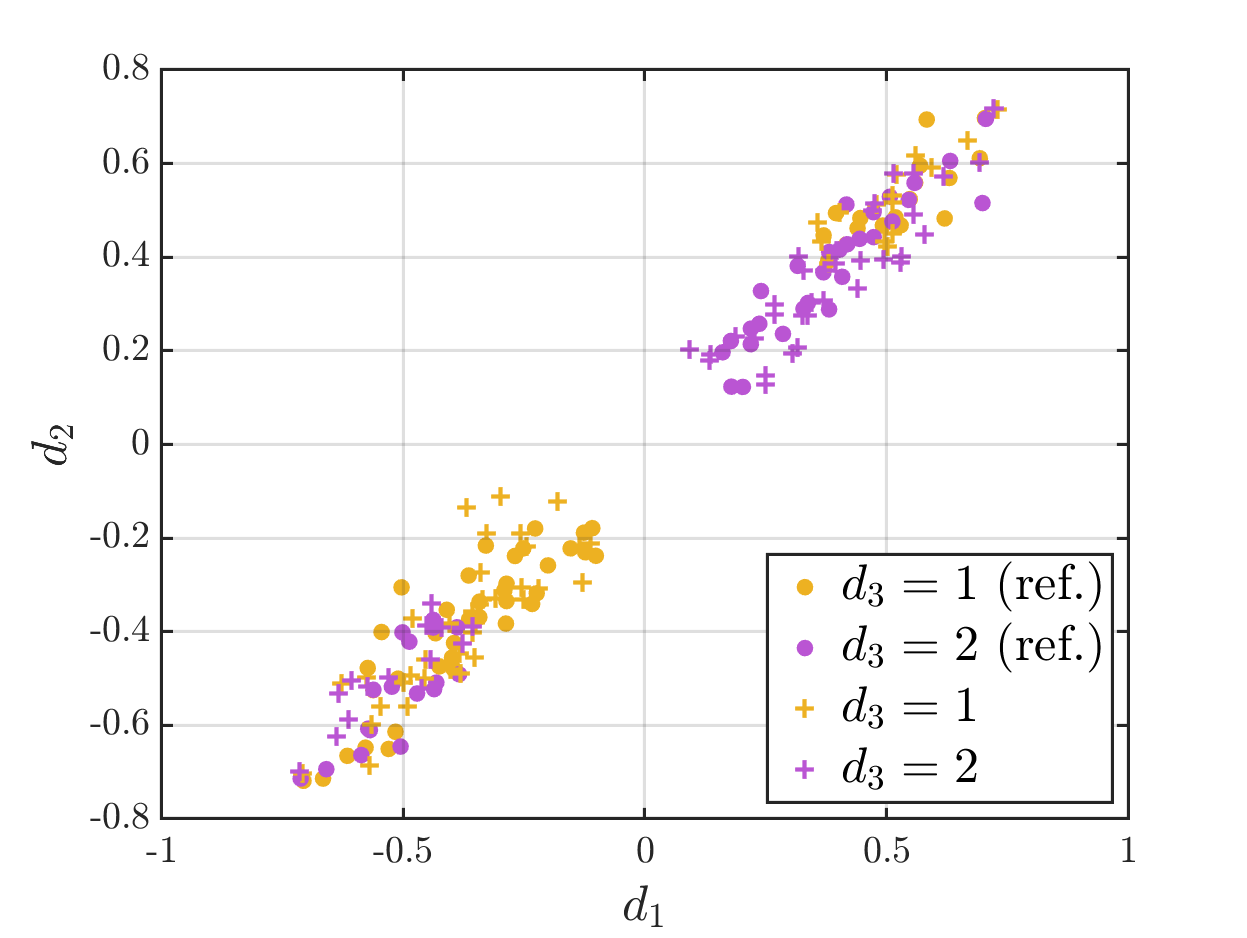}}%
	\caption{Example 2: Comparison of the Pareto fronts and sets for the results with median relative hypervolume error and the reference solution.}
	\label{fig:Example2:Pareto}
\end{figure}

Finally, we show in Figure~\ref{fig:Example2:Pareto} the Pareto front and sets obtained for the median solution at the threshold $\bar{\eta}_q = 0.03$. The two Pareto fronts coincide in the objective space, showing good convergence of the algorithm. Similarly, the Pareto sets of the reference and approximated solutions overlap in the input space. 

\subsection{Example 3: Application to building renovation}
This third example deals with building renovation, which is actually the application that has motivated this work. Because buildings are responsible for $40 \%$ of energy consumption and $36 \%$ of greenhouse gas emissions from energy in Europe, the European union has recently pledged to renovate $35$ million buildings in the next $10$ years \citep{EC2020_24102020}. Building renovation is indeed an important lever since current buildings are not energy-efficient but yet are expected for the most part to still stand by 2050. 

Renovation thus needs to take into account the entire life cycle of the building, which may span over decades. This implies accounting for various uncertainties, be it in socio-economic and environmental conditions or in the variability of the parameters of the selected renovation strategies. This can be achieved using life cycle analysis where two quantities of interest are often considered: the \emph{life cycle cost} ($LCC$) and the \emph{life cycle environmental impact} ($LCEI$). The former includes various costs such as the cost of production of new materials, the related cost of replacement or repair, the labor cost, etc. The latter refers to the overall greenhouse gas emissions over the entire life cycle of the building posterior to the renovation.

The stakeholders need to weigh these two quantities to decide which are the optimal renovation strategies for a given building while accounting for the various sources of uncertainty. To this aim, robust multi-objective optimization may be used as a reliable way of exploring the extremely large design space (\emph{i.e.}, the combination of all possible choices available to the stakeholders). Using $\mathfrak{c}_1 = LCC$ and $\mathfrak{c}_2 = LCEI$, the problem may be formulated as in Eq.~\eqref{eq:RMO_deter} and the proposed methodology may be used to solve it.

\begin{figure}[!ht]
	\centering
	\subfloat[Building facade]{\label{fig:Example3a}\includegraphics[width=0.39\textwidth]{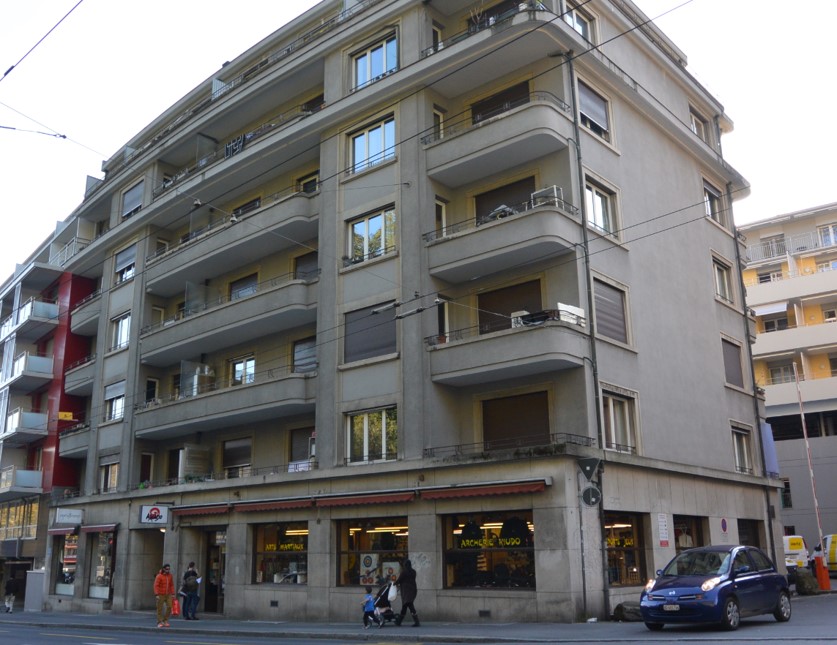}}%
	\subfloat[Possible renovation scenarios]{\label{fig:Example3:b}\includegraphics[width=0.45\textwidth]{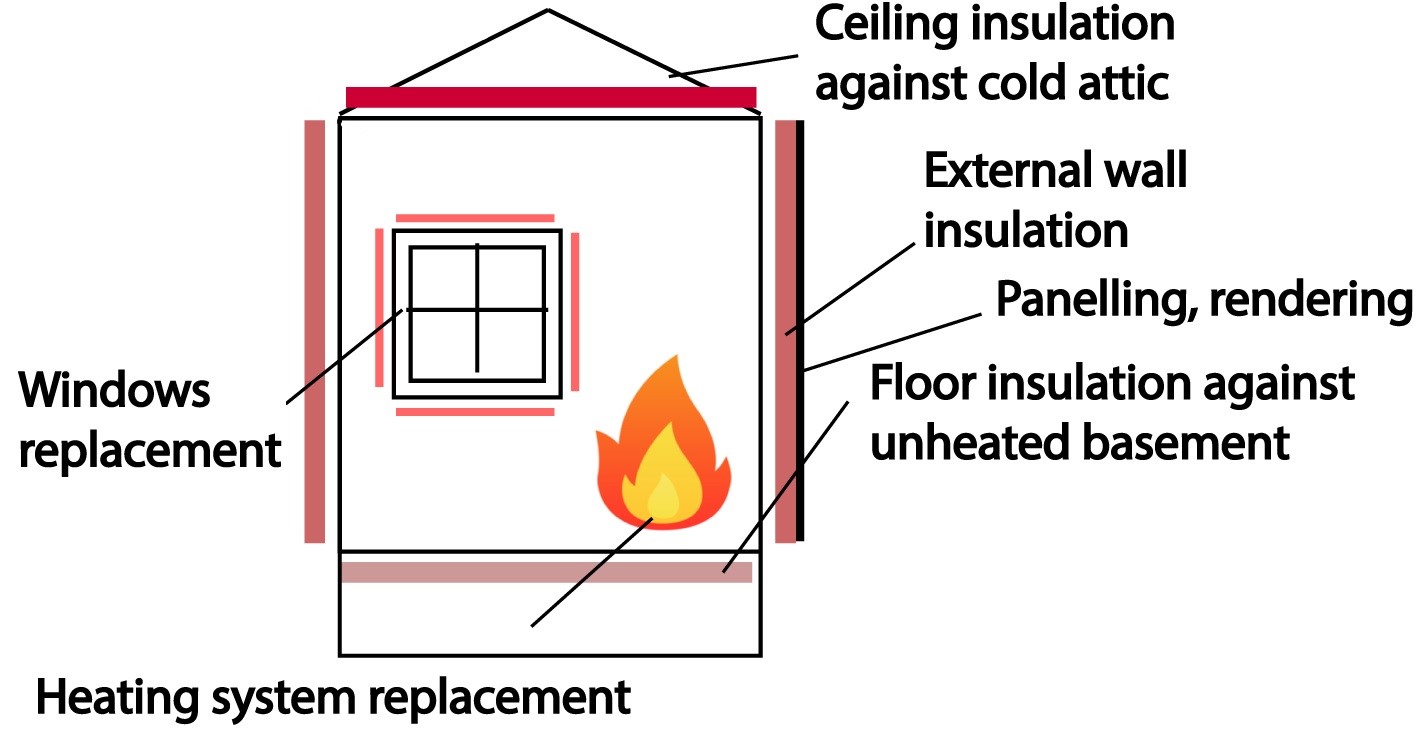}}%
	\caption{Building considered from renovation together with a few possible renovation scenarios (adapted from \citet{GalimshinaBE2020}).}
	\label{fig:Example3}
\end{figure}

As an application, we consider in this paper, a building situated in Western Switzerland and constructed in 1911 (See Figure~\ref{fig:Example3}). The LCA is carried out using a model developed in \citet{GalimshinaBE2020}. The computational model is implemented in Python and a single run lasts a few seconds. The model contains more than a hundred parameters. However, using expert knowledge and global sensitivity analysis, screening allowed us to reduce the input to $23$ parameters, among which $6$ are design parameters and $13$ are environmental variables \citep{GalimshinaBE2020}. The design parameters include $4$ categorical variables as shown in Table~\ref{tab:Example3:Param}, which lead to $3,600$ levels. This includes the $6$ types of heating system: oil, gas, heat pump, wood pellets, electricity and district heating. Various types of walls and windows with different characteristics that are selected from a publicly available catalog are also considered. The remaining two design parameters are the insulation thickness associated to the selected walls and slabs. The environmental variables are all random and their distributions are shown in Table~\ref{tab:Example3:EnvironParam}. They are split into three groups which pertain to the occupancy usage, economic parameters and renovation components variability. 
\begin{table}[!ht]
	\centering
	\caption{Design parameters selected for the building renovation application. Curly brackets $\acc{\cdot}$ correspond to categorical variables. Square brackets $\bra{\cdot}$ define the interval of interest for continuous variables.}
	\label{tab:Example3:Param}
	\setlength{\tabcolsep}{0em}
	\begin{tabular}{lc}
		\hline
		Parameter &    Range \\ \hline
		Heating system &  $\acc{1, \, 2, \enum 6}$   \\
		Exterior wall type  &  $\acc{1, \, 2 \enum 10}$\\
		Exterior wall insulation thickness (m)    &  $\bra{0.1, \,0.5}$             \\ 
		Windows type  &  $\acc{1, \, 2 \enum 10}$   \\
		Slab against unheated type  &  $\acc{1, \, 2 \enum 6}$  \\
		Slab against unheated insulation thickness (m)    &  $\bra{0.1, \,0.5}$             \\ \hline
	\end{tabular}			
\end{table}
\begin{table*}[!ht]
	\centering
	\caption{Distribution of the environmental variables for the building renovation application.}
	\label{tab:Example3:EnvironParam}
	\begin{threeparttable} 
		\begin{tabular}{lccc}
			\toprule
			Parameter &    Distribution & Parameter 1\tnote{a} & Parameter 2\tnote{b}\\ \hline
			Room temperature  (degree C)  &  Uniform   &  ${ 20}$ &           ${23}$          \\
			Airflow (m$^3/$h,m$^2$)    &  Uniform   & ${0.7}$  &		${1}$            \\ \hline
			Inflation rate  ($\%$)   &  Uniform   & ${0.005}$  &           ${0.02}$           \\ 
			Discount rate ($\%$)   &  Uniform   &  ${0.035}$ &           ${0.045}$          \\ \hline
			Uncertainty on exterior wall cost ($\%$)  &  Uniform   & ${ -20}$ &     ${ 20}$          \\ 
			Uncertainty on exterior wall GWP\tnote{c} ($\%$)  &  Uniform   & ${ -20}$ &     ${ 20}$          \\
			Uncertainty on windows cost ($\%$)  &  Uniform   & ${ -20}$ &     ${20}$          \\
			Uncertainty on windows GWP ($\%$)  &  Uniform   & ${ -20}$ &     ${20}$          \\
			Uncertainty on slab against unheated cost ($\%$)  &  Uniform   & ${ -20}$ &     ${ 20}$          \\
			Uncertainty on slab against unheated GWP ($\%$)  &  Uniform   & ${ -20}$ &     ${ 20}$          \\
			Uncertainty on slab against unheated RSL\tnote{d} ($\%$)  &  Uniform   & ${ -20}$ &     ${ 20}$          \\ 
			Heating efficiency loss ($\%$) & Uniform  & ${15}$ &     ${25}$           \\
			Exterior wall conductivity   &  Gumbel   & ${ 0.69}$ &     ${ 0.18}$          \\
			Exterior wall RSL   (years)   &  Lognormal   & ${ 32}$ &     ${ 2}$          \\ 
			Windows RSL   (years)   &  Lognormal   & ${ 32}$ &     ${ 2}$          \\ 
			\bottomrule
		\end{tabular}
		{\footnotesize	
			\begin{tablenotes}
				\item[a] Corresponds to the lower bound for uniform distributions and to the mean for non uniform distributions.
				\item[b] Corresponds to the upper bound for uniform distributions and to the standard deviation for non uniform distributions.
				\item[c] GWP stands for global warming potential.
				\item[d] RSL stands for reference service life.
			\end{tablenotes}
		}
	\end{threeparttable}
\end{table*}

\begin{figure}[!ht]
	\centering
	\subfloat[Convergence for $\mathfrak{c}_1 = LCC$]{\label{fig:Example3:Conv:a}\includegraphics[width=0.45\textwidth]{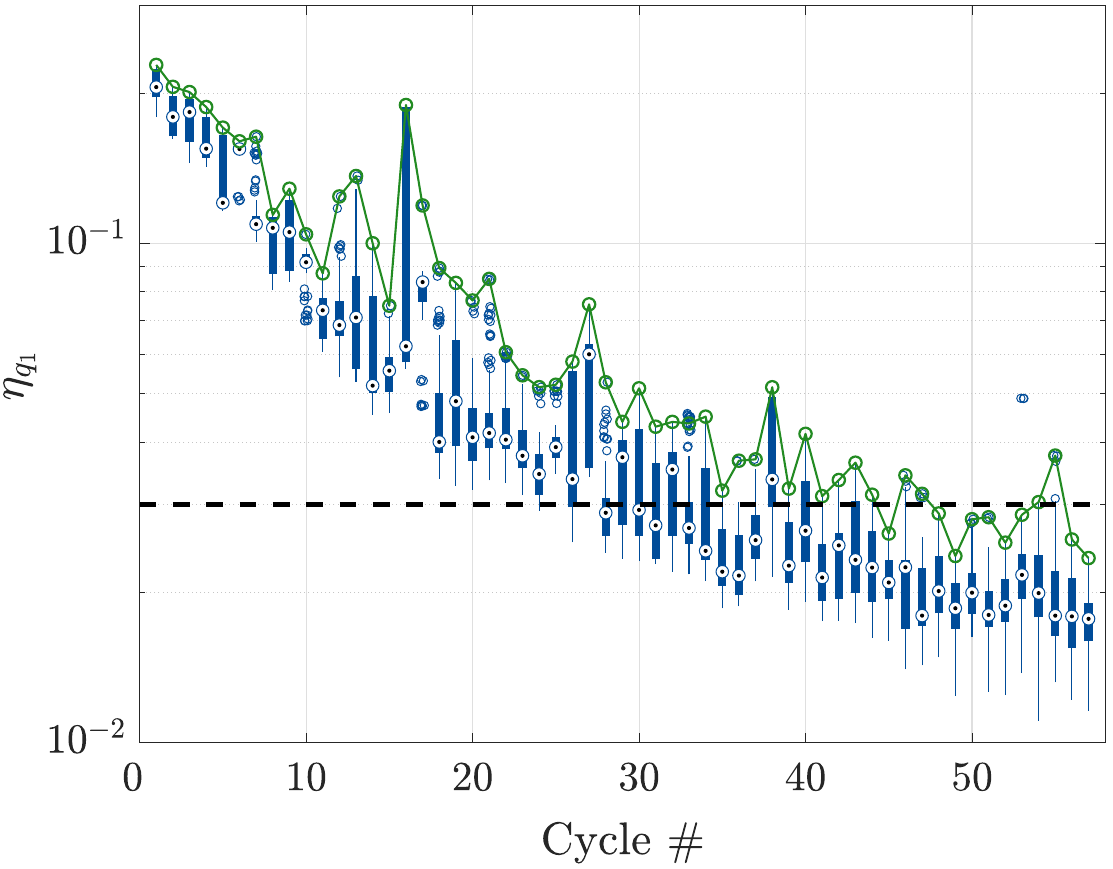}}%
	\subfloat[Convergence for $\mathfrak{c}_2= LCEI$]{\label{fig:Example3:Conv:b}\includegraphics[width=0.45\textwidth]{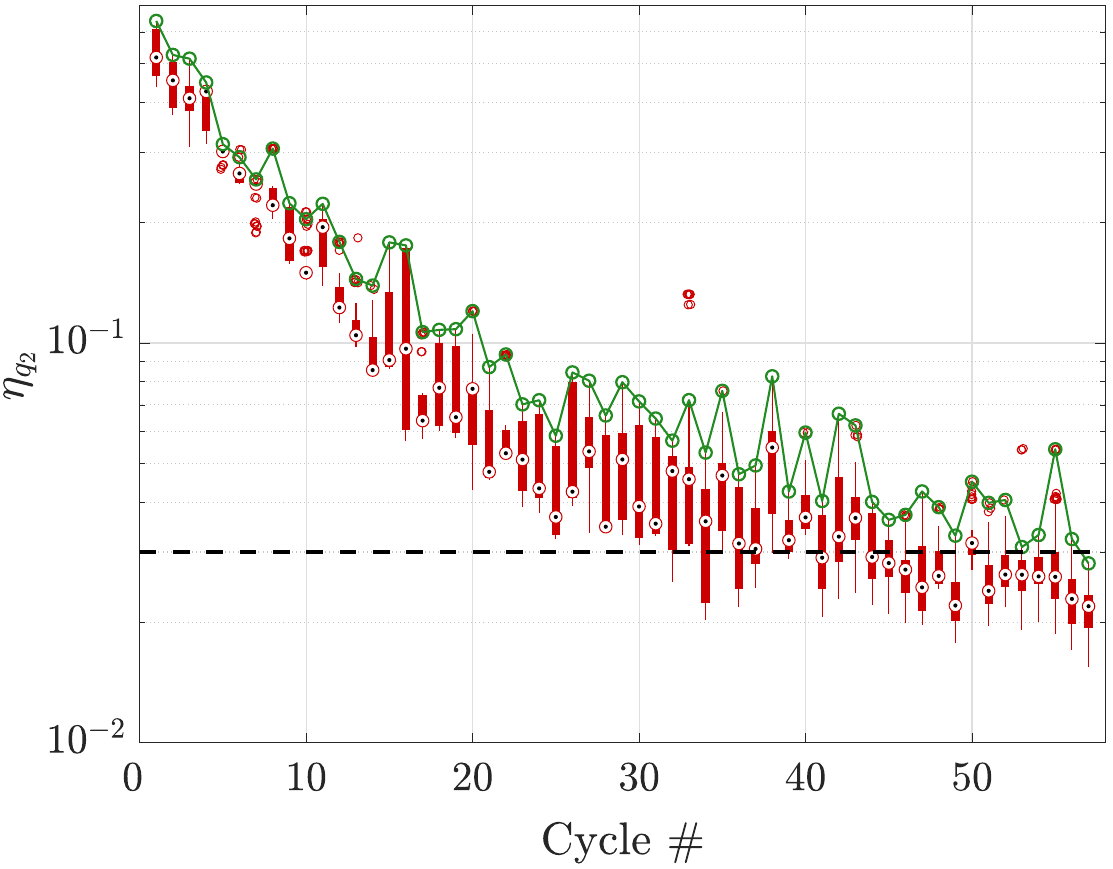}}%
	\caption{Example 3: Relative quantile error of the entire Pareto front at the end of each cycle.}
	\label{fig:Example3:Conv}
\end{figure}

\begin{figure}[!ht]
	\centering
	\includegraphics[width=0.50\textwidth]{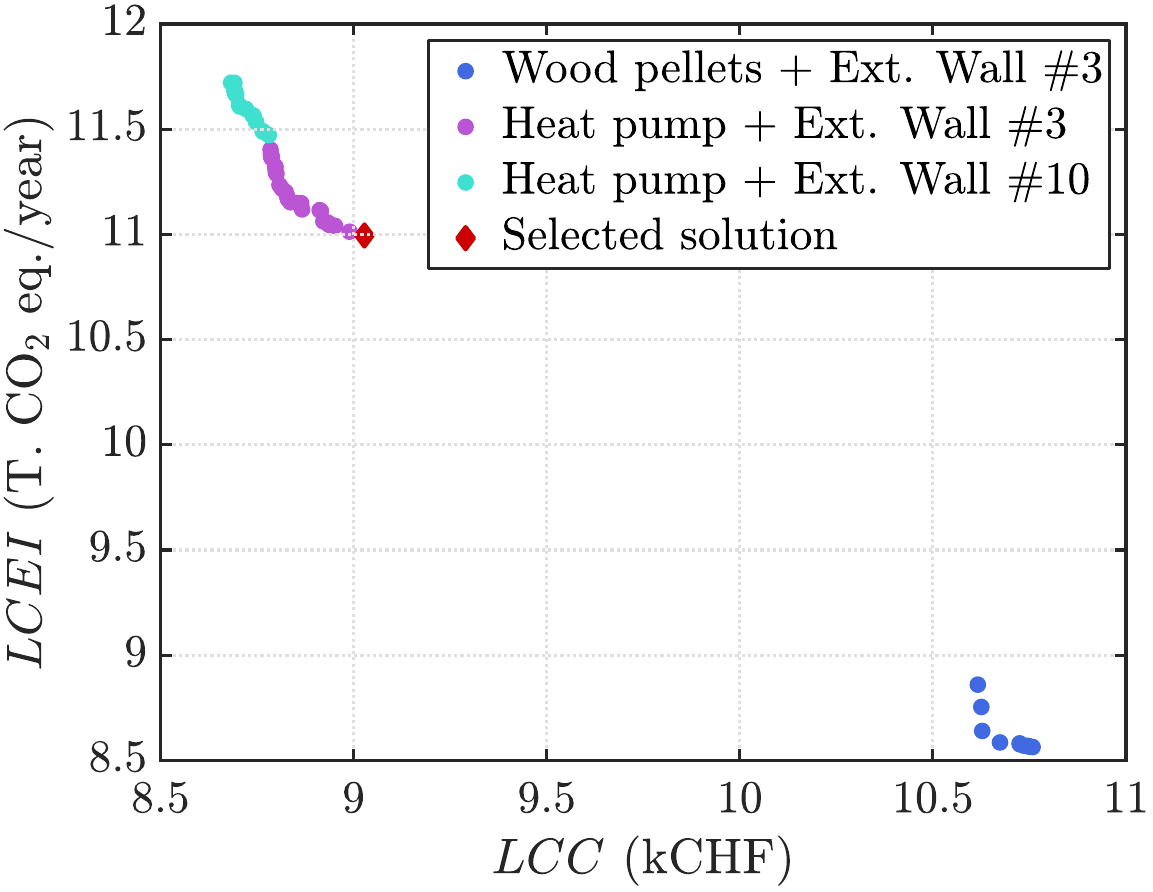}%
	\caption{Example 3: Pareto front and selected solution.}
	\label{fig:Example3:Pareto}
\end{figure}

The analysis is initialized with an experimental design of size $n_0 = 3 M = 69$ points drawn using an optimized Latin hypercube sampling scheme. The threshold for convergence is set to $\bar{\eta}_q = 0.03$, which resulted in $57$ cycles of the algorithm as shown in Figure~\ref{fig:Example3:Conv}. This corresponds to a total of $271$ model evaluations. The Pareto front is shown in Figure~\ref{fig:Example3:Pareto}. It features some discontinuities which are due to some noticeable changes in the properties of the categorical variables. The main driver to decrease both $LCC$ and $LCEI$ is the heating system. The upper part of the Pareto front corresponds to a heat pump which leads to small values of $LCC$ and larger values of $LCEI$. This is in contrast with the wood pellets which correspond to the lower part of the Pareto front.  For this example, we eventually select one solution, which is in the upper part and is highlighted by the red diamond in Figure~\ref{fig:Example3:Pareto}. This choice reflects a preference on the cost with respect to the environmental impact. Table~\ref{tab:Example3:Result} shows the detailed values of this selected renovation strategy.
\begin{table}[!ht]
	\centering
	\caption{Selected renovation strategy from the Pareto front for the building renovation.}
	\label{tab:Example3:Result}
	\setlength{\tabcolsep}{0em}
	\begin{tabular}{lc}
		\hline
		Parameter &    Value \\ \hline
		Heating system &  Heat pump   \\
		Exterior wall type  &  $\#4$ \\
		Exterior wall insulation thickness (m)    &  $0.4778$             \\ 
		Windows type  & $\#9$  \\
		Slab against unheated type  & $\#5$  \\
		Slab against unheated insulation thickness (m)    &  $0.4070$             \\ \hline
	\end{tabular}			
\end{table}

\begin{figure}[!ht]
		\centering
		\subfloat[$LCC$]{\label{fig:Example3:Y_Y:a}\includegraphics[width=0.45\textwidth]{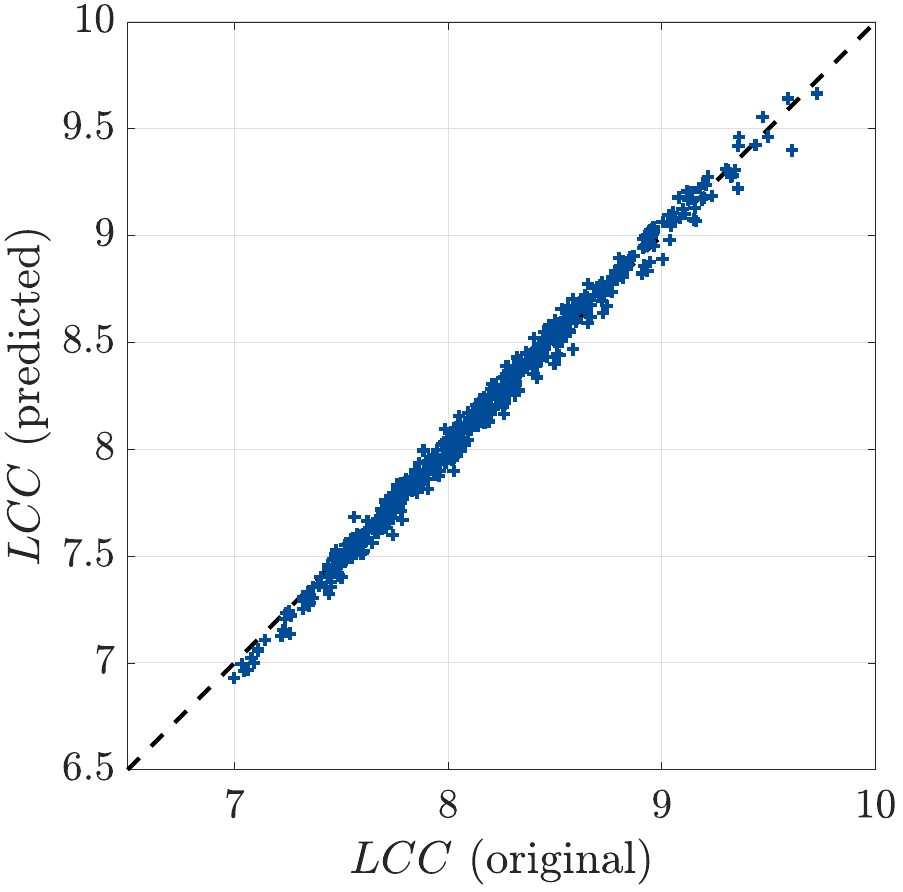}}%
		\subfloat[$LCEI$]{\label{fig:Example3:Y_Y:b}\includegraphics[width=0.45\textwidth]{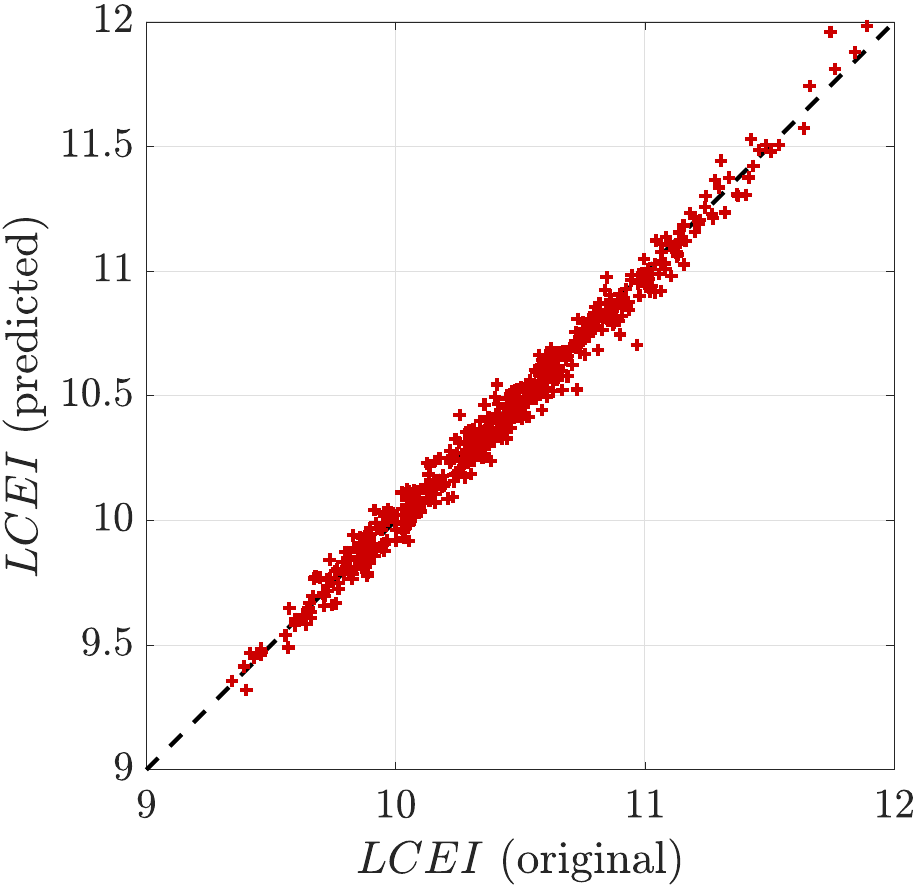}}%
		\caption{Example 3: Original \emph{vs.} predicted responses for a subset of the Monte Carlo set used to compute the quantiles at the selected solution.}
		\label{fig:Example3:Y_Y}
	\end{figure}

Finally, to assess the accuracy of this analysis, we sample a validation set \\ $\mathcal{C}_{\textrm{val}} = \acc{\prt{\ve{d}^\ast, \ve{z}^{(i)}}, i = 1 \enum 500}$, where $\ve{d}^\ast$ is the selected solution. Figure~\ref{fig:Example3:Y_Y} compares the prediction by the final surrogate model and the original response on this validation set. As expected, the surrogate model is able to correctly approximate the original model around the chosen solution. This is confirmed by comparing the normalized mean-square error (NMSE) and the $90\%$-quantile shown in Table~\ref{tab:Example3:NMSE_Q}. Even though the Monte Carlo sample set size is reduced, the estimated quantiles allow us to have an idea of how accurate the surrogate models are. More specifically the relative errors of the quantiles of $LCC$ and $LCEI$ due to the introduction of the surrogates are approximately $0.2\%$ and $0.4\%$, respectively. The NMSE for $LCEI$ is slightly larger compared to that of $LCC$, which is consistent with the convergence history in Figure~\ref{fig:Example3:Conv}. This value could be reduced by selecting a smaller value of $\bar{\eta}_q$. This would however lead to a larger computational cost and analysis run time.
	\begin{table}[!ht]
		\centering
		\caption{Validation of the selected solution.}
		\label{tab:Example3:NMSE_Q}
		\setlength{\tabcolsep}{0.3em}
		\begin{tabular}{lcccc}
			\hline
			& \multicolumn{2}{c}{$LCC$ (kCHF)} & \multicolumn{2}{c}{$LCEI$ (T. CO$_{2}$ eq. / a)} \\
			Parameter &    Original & Surrogate & Original & Surrogate \\ \hline
			Quantile &  $8.9386$ & $8.9530$ & $11.0599$ & $11.0183$   \\
			NMSE  &  & $0.0095$ & & $0.0146$   \\ \hline
		\end{tabular}			
	\end{table}

\section{Conclusion}
In this paper, we proposed a methodology for the solution of multi-objective robust optimization problems involving categorical and continuous variables. The problem was formulated using quantiles as a measure of robustness, the level of which can be set to control the desired degree of robustness of the solution. A nested solution scheme was devised, where optimization is carried out in the outer loop while uncertainty propagation is performed in the inner loop. This however results in a stochastic optimization problem which may be cumbersome to solve. The concept of common random numbers was introduced to approximate this stochastic problem by a deterministic one, which is much easier to solve. This allows us then to use a general-purpose multi-objective solver, namely the non-dominated sorting genetic algorithm II (NSGA-II). 

To reduce the computational burden of this nested solution scheme, Kriging was introduced in the proposed framework using an adaptive sampling scheme. This is enabled by checking the accuracy of the quantile estimates in areas of interest, namely the area around the Pareto front. Finally, the proposed approach was adapted to account for mixed categorical-continuous parameters. Two validation examples were built analytically while considering all the characteristics of the problem at hand. 

The methodology was then applied to the optimization of renovation scenarios for a building considering uncertainties in its entire life cycle post-renovation. This has shown that a driving parameter for such a renovation is the replacement of the heating system by either wood pellets or heat pump. This result was validated by running the original model around the selected solution, hence showing that the final surrogate model was accurate. This methodology was eventually applied in a detailed analysis presented in \citet{GalimshinaENB2021}.

\section{Competing interests}
The authors declare that they have no known competing financial interests or personal relationships that could have appeared to influence the work reported in this paper.

\section{Replication of results}
The results presented in this paper were obtained using an implementation of the proposed methodology within \textsc{UQLab} (www.uqlab.com) \citep{MarelliUQLab2014}, a \textsc{Matlab}-based framework for uncertainty quantification. The codes can be made available upon request.


\bibliographystyle{chicago}
\bibliography{Biblio_Moustapha_etal_2022_SAMO}

\appendix

\section{Non-dominated sorting genetic algorithm II}\label{app:NSGA}
Generally speaking, genetic algorithms \citep{Goldberg1989} are a sub-class of evolutionary methods where a population of individuals is evolved so as to produce better and better solutions until convergence. Genetic algorithms attempt to mimic natural evolution theory where individuals are combined through so-called \emph{cross-over} and \emph{mutation} operators, the idea being that the fittest individuals would pass their genes to the next generation. The non-dominated sorting genetic algorithm II (NSGA-II) belongs to a class of algorithms developed specifically to handle multi-objective optimization problems. The latter introduce a new approach for assigning fitness according to a rank derived from a non-dominated sorting. NSGA-II is an elitist and fast non-dominated sorting algorithm introduced by \citet{Deb2002} as an enhancement of the original NSGA \citep{Deb1999}.  Even though NSGA has shown to be a popular evolutionary method, it features a computationally complex and slow sorting algorithm and, most of all, lacks elitism. Elitism, a property by which the fittest solutions in a given generation are copied to the next one so as not to lose the best genes, has indeed proved to accelerate convergence \citep{Zitzler2000}. The general workflow of NSGA-II is illustrated in Figure~\ref{fig:FlowchartNSGA}. The following section details the important steps of NSGA-II as described in Algorithm~\ref{alg:nsga2}:
\begin{figure}[!ht]
	\centering
	\includegraphics[width=0.5\textwidth]{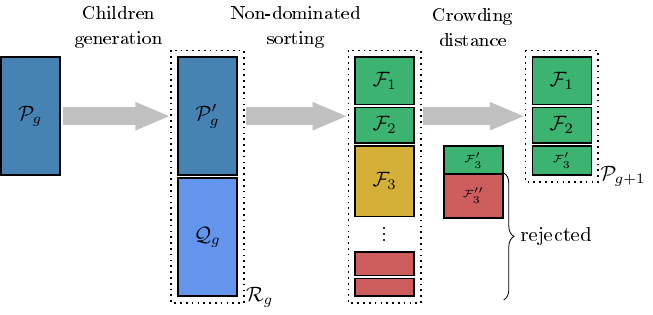}%
	\caption{Flowchart of NSGA-II (adapted from \citet{Deb2002}).}
	\label{fig:FlowchartNSGA}
\end{figure} 

\begin{algorithm*}[!ht]
	\caption{Non-dominated sorting genetic algorithm II \citep{Deb2002}}
	\begin{algorithmic}[1]
		\Require{}
		\Statex Sample initial population $\mathcal{P}_0$ of size $L$		\Comment{\color{DarkBlue} {\scriptsize{\emph{e.g.}, $L = 100$ }} \color{black}}
		\Statex Set maximum number of iterations $G_{\textrm{max}}$		\Comment{\color{DarkBlue} {\scriptsize{\emph{e.g.}, $G_{\textrm{max}} = 100$ }} \color{black}}
		\Statex Set $g = 0$; \texttt{NotConverged} $= \texttt{true}$,
		\Statex \hrulefill
		\While{\texttt{NotConverged} $== \texttt{true}$}
		\State Perform tournament selection: $\mathcal{P}_g^\prime \leftarrow \mathcal{P}_g$ \Comment{\color{DarkGreen} {\scriptsize{$\mathcal{P}_g^\prime$ is also of size $L$ }} \color{black}}
		\State Generate offsprings $Q_g$ \Comment{\color{DarkGreen} {\scriptsize{using cross-over and mutation operators}} \color{black}}
		\State Set $\mathcal{R}_g = \mathcal{P}_g^\prime \cup \mathcal{Q}_g$ \Comment{\color{DarkGreen} {\scriptsize{Combine parents and offspring }} \color{black}}
		\State Perform non-dominated sorting $\mathcal{F} = \acc{\mathcal{F}_1, \mathcal{F}_2, ...}$
		\State Set $\mathcal{P}_{g+1} = \emptyset, \quad i = 1$
		\While{$ \textrm{Card}\prt{\mathcal{P}_{g+1}} < L$}
		\State $\mathcal{P}_{g+1} = \mathcal{P}_{g+1} \cup \mathcal{F}_i$
		\State $i \leftarrow i+1$
		\EndWhile
		\State Sort $\mathcal{F}_i$ \Comment{\color{DarkGreen} {\scriptsize{w.r.t. the individuals crowding distance}} \color{black}}
		\State Set $\mathcal{F}_i^\prime = $ \texttt{sorted}$\prt{\mathcal{F}_i}\bra{1:L-\textrm{Card}\prt{\mathcal{P}_{g+1}}}$  \Comment{\color{DarkGreen} {\scriptsize{Select the $L-\textrm{Card}\prt{\mathcal{P}_{g+1}}$ best elements in $\mathcal{F}_i$  }} \color{black}}
		\State Set $\mathcal{P}_{g+1} = \mathcal{P}_{g+1} \cup \mathcal{F}_i^\prime$
		\Let{$g$}{$g+1$}
		\If{Convergence is achieved or $g = G_{\textrm{max}}$} \State \texttt{NotConverged} $= \texttt{false}$  \EndIf
		\EndWhile
	\end{algorithmic}
	\label{alg:nsga2}
\end{algorithm*}

\begin{enumerate}
	\item \textbf{Children generation:} Following the initialization where the population $\mathcal{P}_g$ is sampled, the first step in Algorithm~\ref{alg:nsga2} is binary tournament selection, which is used to pre-select good individuals for reproduction. The idea is to form a new population $\mathcal{P}^\prime_g$ by selecting the best among two (or more) randomly selected individuals. The new population is then of the same size as the original but comprises repetitions. Individuals in this population constitute the basis for the generation of children using cross-over and mutation operators. The former consists in combining two parents to generate two offsprings. There exists a wide variety of cross-over operators \citep{Umbarkar2015} and NSGA-II uses the so-called \emph{simulated binary cross-over} (SBX). Mutation is used to prevent premature convergence to a local minimum. Following a predefined probability, a new offspring is mutated, \emph{i.e.}, some of its components are randomly perturbed, thus allowing for more exploration of the design space. The \emph{polynomial mutation} operator has been developed for NSGA-II. The resulting set of offsprings is denoted by $\mathcal{Q}_g$.
	\item \textbf{Non-dominated sorting:} Once the new population $\mathcal{R}_g$ is generated by assembling $\mathcal{P}^\prime_g$ and $\mathcal{Q}_g$, it is subjected to a recursive non-dominated sorting. This means that the set of all non-dominated individuals are identified and assigned rank one. This set, denoted by $\mathcal{F}_1$, is then set aside and non-dominated sorting is again performed on $\mathcal{R}_g \setminus \mathcal{F}_1$ to find the front of rank two $\mathcal{F}_2$. The process is repeated until all individuals have been assigned a rank. The population for the next generation is then composed of the individuals of the first ranks whose cumulated cardinality is smaller than $L$.
	\item \textbf{Crowding distance assignment:}  To reach $L$ offsprings, it is necessary to select only a subset of the population in the following front. To do so, the crowding distance, which measures the distance of a given solution to the closest individuals in its rank, is used. Diversity in the Pareto front is hence enforced by choosing for the same rank solutions with the largest crowding distance. Once the remaining solutions are chosen according to their crowding distance ranking, the new population $\mathcal{P}_{g+1}$ is defined. 
\end{enumerate}
This entire process is repeated until convergence is reached.

\end{document}